\documentclass[twocolumn,prb,superscriptaddress]{revtex4-2}
\pdfoutput=1
\usepackage{graphicx}
\usepackage{amssymb,amsmath}
\usepackage{bm}
\usepackage{dcolumn}
\usepackage{subfigure}
\usepackage{float}
\usepackage[OT1]{fontenc} 
\usepackage{url}
\usepackage{slashed}
\usepackage{color}
\usepackage{verbatim}
\usepackage{enumitem}
\usepackage{txfonts}
\usepackage{soul,physics}
\usepackage[driverfallback=dvipdfm]{hyperref}
\hypersetup{pdfpagemode=FullScreen,colorlinks=true,breaklinks,urlcolor=blue,linkcolor=blue,citecolor=blue}

\usepackage{subfigure}
\usepackage{comment}
\usepackage{hyperref}
\usepackage{amssymb}
\usepackage{bm}
\usepackage{graphicx}
\usepackage{amsmath,amssymb}
\usepackage{tikz,fp}
\usepackage{tikz-cd}
\usetikzlibrary{arrows}
\usetikzlibrary{intersections}
\usetikzlibrary{shapes.geometric}
\usetikzlibrary{decorations.pathmorphing, patterns,shapes,fixedpointarithmetic}
\usetikzlibrary{decorations.markings}

\pgfdeclarepatternformonly{south west lines}{\pgfqpoint{-0pt}{-0pt}}{\pgfqpoint{3pt}{3pt}}{\pgfqpoint{3pt}{3pt}}{
	\pgfsetlinewidth{0.4pt}
	\pgfpathmoveto{\pgfqpoint{0pt}{0pt}}
	\pgfpathlineto{\pgfqpoint{3pt}{3pt}}
	\pgfpathmoveto{\pgfqpoint{2.8pt}{-.2pt}}
	\pgfpathlineto{\pgfqpoint{3.2pt}{.2pt}}
	\pgfpathmoveto{\pgfqpoint{-.2pt}{2.8pt}}
	\pgfpathlineto{\pgfqpoint{.2pt}{3.2pt}}
	\pgfusepath{stroke}}

\tikzset{
	mid arrow/.style={postaction={decorate,decoration={
				markings,
				mark=at position .575 with {\arrow{stealth}}
	}}},
	near arrow/.style={postaction={decorate,decoration={
				markings,
				mark=at position .275 with {\arrow{stealth}}
	}}},
	far arrow/.style={postaction={decorate,decoration={
				markings,
				mark=at position .800 with {\arrow{stealth}}
	}}},
	snake arrow/.style={fixed point arithmetic, decorate, decoration={snake,amplitude=2pt, segment length=11pt},postaction={decoration={markings,mark=at position 0.625 with {\arrow{stealth}}},decorate}},
}
\tikzset{
  baseline = -0.5ex,
  wavy/.style = {
    thick,
    decorate,
    decoration={snake,amplitude=2pt,segment length=5pt}},
  sdot/.style = {
    circle,
    draw=none,
    fill=black,
    minimum size=2.5pt,
    inner sep=0pt},
  bdot/.style = {
    circle,
    draw=none,
    fill=black,
    minimum size=4pt,
    inner sep=0pt},
  svertex/.style = {
    circle,
    draw=black,
    thick,
    fill=lightgray,
    minimum size=8pt,
    inner sep=1pt},
  bvertex/.style = {
    circle,
    draw=black,
    thick,
    fill=lightgray,
    minimum size=24pt},
  bvertexsmall/.style = {
    circle,
    draw=black,
    thick,
    fill=lightgray,
    minimum size=7pt},
  bvertexnormal/.style = {
    circle,
    draw=black,
    thick,
    fill=lightgray,
    minimum size=16pt},
  dvertex/.style = {
    circle,
    draw=black,
    thick,
    fill=gray,
    minimum size=25pt}}

\newcommand{\iu}{{i\mkern1mu}}
\newcommand*\diff{\mathop{}\!\mathrm{d}}

\begin{document}
	
	\title{ Full Counting Statistics across the Entanglement Phase Transition of\\ Non-Hermitian Hamiltonians with Charge Conservations}
	\author{Tian-Gang Zhou}
    \affiliation{Institute for Advanced Study, Tsinghua University, Beijing, 100084, China}

   \author{Yi-Neng Zhou}
    \affiliation{Institute for Advanced Study, Tsinghua University, Beijing, 100084, China}

	\author{Pengfei Zhang}
	\thanks{pengfeizhang.physics@gmail.com}
	\affiliation{Department of Physics, Fudan University, Shanghai, 200438, China}
    \affiliation{Shanghai Qi Zhi Institute, AI Tower, Xuhui District, Shanghai 200232, China}
	\date{\today}

	\begin{abstract}
    Performing quantum measurements produces not only the expectation value of a physical observable $O$ but also the probability distribution $P(o)$ of all possible outcomes $o$. The full counting statistics (FCS) $Z(\phi, O)\equiv \sum_o e^{i\phi o}P(o)$, a Fourier transform of this distribution, contains the complete information of the measurement outcome. In this work, we study the FCS of $Q_A$, the charge operator in subsystem $A$, for 1D systems described by non-Hermitian SYK-like models, which are solvable in the large-$N$ limit. In both the volume-law entangled phase for interacting systems and the critical phase for non-interacting systems, the conformal symmetry emerges, which gives $F(\phi, Q_A)\equiv \log Z(\phi, Q_A)\sim \phi^2\log |A|$. In short-range entangled phases, the FCS shows area-law behavior which can be approximated as $F(\phi, Q_A)\sim (1-\cos\phi) |\partial A|$ for $\zeta \gg J$, regardless of the presence of interactions. Our results suggest the FCS is a universal probe of entanglement phase transitions in non-Hermitian systems with conserved charges, which does not require the introduction of multiple replicas. We also discuss the consequence of discrete symmetry, long-range hopping, and generalizations to higher dimensions.
	\end{abstract}

 \maketitle

     \section{Introduction}
     Unitary evolution and projective measurements are basic building blocks of quantum operations. Recent studies unveil novel entanglement phase transitions driven by their competition in systems under repeated measurements \cite{mazzucchiQuantumMeasurementinducedDynamics2016,liQuantumZenoEffect2018,skinnerMeasurementInducedPhaseTransitions2019,liMeasurementdrivenEntanglementTransition2019,szyniszewskiEntanglementTransitionVariablestrength2019,chanUnitaryprojectiveEntanglementDynamics2019,vasseurEntanglementTransitionsHolographic2019,zhouEmergentStatisticalMechanics2019,gullansScalableProbesMeasurementInduced2020,jianMeasurementinducedCriticalityRandom2020,fujiMeasurementinducedQuantumCriticality2020,zabaloCriticalPropertiesMeasurementinduced2020,gullansDynamicalPurificationPhase2020,choiQuantumErrorCorrection2020,baoTheoryPhaseTransition2020,nahumMeasurementEntanglementPhase2021,PhysRevB.103.174309,sangMeasurementprotectedQuantumPhases2021,albertonEntanglementTransitionMonitored2021,lavasaniMeasurementinducedTopologicalEntanglement2021,PhysRevB.103.224210,LeGal:2022rwf}. For small/large measurement rates, the steady state in a typical quantum trajectory is volume-law/area-law entangled. Later, it is realized that entanglement transitions also exist in general non-unitary dynamics. In particular, Sachdev-Ye-Kitaev (SYK) large-$N$ solvable models with non-Hermitian Hamiltonians have been proposed, where the transition of the 2nd R\'enyi entropy is mapped to a transition of classical spins \cite{jianMeasurementInducedPhaseTransition2021a,zhangUniversalEntanglementTransitions2022,liuNonunitaryDynamicsSachdevYeKitaev2021,zhang2021emergent,zhang2022quantum,PhysRevB.106.224305}. The corresponding order parameter is the quantum correlation $G^{ud}$ between forward/backward evolution branches on the (replicated) Keldysh contour: the R\'enyi entropy can be expressed as a correlator of the replica twist operator $\mathcal{T}_r$. For interacting systems, the spin model is $Z_4$ symmetric. When $G^{ud}\neq 0$, the spin model is in the ordered phase. The insertion of $\mathcal{T}_r$ excites a domain wall. This leads to a volume-law entangled phase. In non-interacting systems, the $Z_4$ symmetry is promoted to an $O(2)$ symmetry. Consequently, the domain wall is replaced by a half-vortex pair, which gives rise to logarithmic entanglement entropy. For $G^{ud}=0$, the spin model is disordered regardless of the presence of interactions, which corresponds to an area-law entangled phase.

    \begin{figure}[tb]
        \centering
        \includegraphics[width=0.9\linewidth]{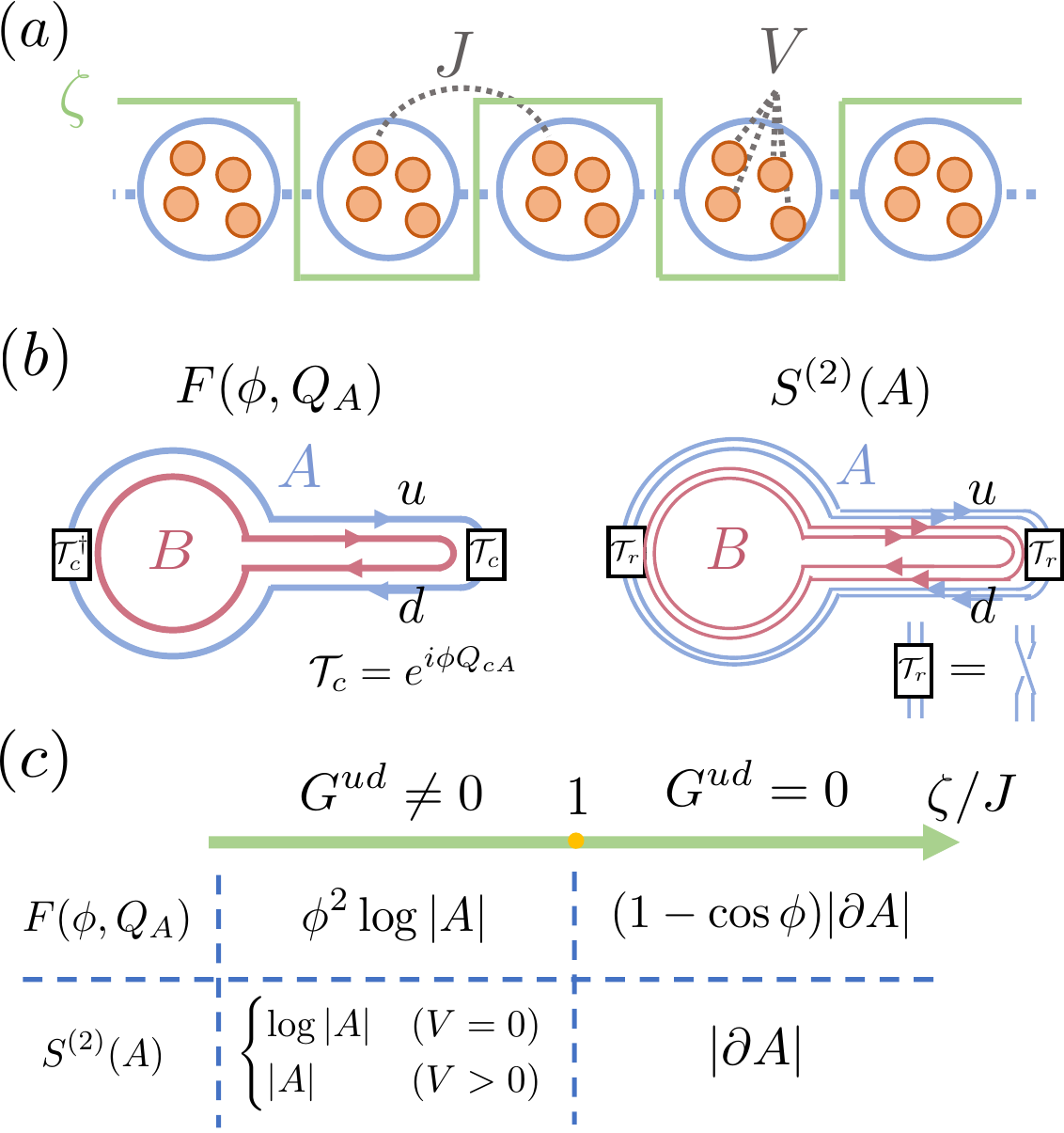}
        \caption{(a). Schematics of the non-Hermitian complex SYK model. $J$ and $V$ are SYK random hopping and on-site interactions. $\zeta$ is a staggered imaginary potential. (b). A comparison between the path integral representations of $F(\phi,Q_A)$ and $S^{(2)}(A)$. Both quantities can be expressed as a correlator of twist operators. (c). The phase diagram of the non-Hermitian complex SYK model. The scaling of $F(\phi,Q_A)$ changes qualitatively when we tune $\zeta/J$ across the entanglement phase transition. }
        \label{fig:schemticas}
    \end{figure}

     On the other hand, it is known that the existence of conserved charges plays an important role in the many-body dynamics of quantum information \cite{rakovszkyDiffusiveHydrodynamicsOutoftimeordered2018,khemaniOperatorSpreadingEmergence2018a,guoTransportChaosLattice2019b,friedmanSpectralStatisticsManyBody2019,rakovszkySubballisticGrowthEnyi2019a,zhouDiffusiveScalingEnyi2020,huangDynamicsRenyiEntanglement2020,kudler-flamInformationScramblingConservation2022,agarwalEmergentSymmetryBrownian2022,zhangInformationScramblingEntanglement2023}. For example, the evolution of the out-of-time-correlator shows a power-law tail due to the charge diffusion in systems with $U(1)$ symmetry \cite{rakovszkyDiffusiveHydrodynamicsOutoftimeordered2018,khemaniOperatorSpreadingEmergence2018a}. In this work, we explore the signature of conserved charges across the entanglement phase transition in large-$N$ non-Hermitian complex SYK chains. We compute the steady-state full counting statistics (FCS) \cite{levitovElectronCountingStatistics1996,klichQuantumNoiseEntanglement2009,songEntanglementEntropyCharge2011,calabreseExactRelationsParticle2012,susstrunkFreeFermionsLine2013,songGeneralRelationEntanglement2010,levineFullCountingStatistics2012a,songBipartiteFluctuationsProbe2012,ivanovCharacterizingCorrelationsFull2013,eislerUniversalityFullCounting2013,eislerFullCountingStatistics2013,klichNoteFullCounting2014,bertiniNonequilibriumFullCounting2022,PotterFieldTheoryCharge2022,mccullochFullCountingStatistics2023,FujiChargeFluctuationChargeresolved2023}
     \begin{equation}\label{eq:defFCS}
     Z(\phi, Q_A)= \lim_{t\rightarrow \infty}~\text{tr}[\rho(t)e^{i\phi Q_A}]\equiv e^{-F(\phi,Q_A)}.
     \end{equation} 
     Here $Q_A=\sum_{x\in A}Q_x$ is the total charge in subsystem $A$, which will be described more precisely later. We choose the convention that $\phi \in(-\pi,\pi]$. {FCS of charge operators are also known as the disorder parameter in \cite{PhysRevB.106.094415,PhysRevB.104.L081109,10.21468/SciPostPhys.13.6.123}.} We mainly focus on initial states described by the thermofield double (TFD) state \cite{Israel:1976ur,maldacenaEternalBlackHoles2003}, which has been widely studied in both high energy and condensed matter physics. The FCS, which takes the form of a generating function, contains the complete information of charge fluctuations in the subsystem $A$. A series of works have observed that charge fluctuations and charge statistics are closely related to entanglement entropy \cite{klichQuantumNoiseEntanglement2009,songEntanglementEntropyCharge2011,calabreseExactRelationsParticle2012,susstrunkFreeFermionsLine2013}. Also, entanglement entropy and FCS work well to characterize bulk and edge spectrum problems \cite{eislerFullCountingStatistics2013,eislerUniversalityFullCounting2013}. Our work provides another perspective to check the profound relation between FCS and entanglement entropy.

     We show that in our setup, FCS can also be viewed as a correlator of twist operators $\mathcal{T}_c$, which now generates a relative phase rotation in the charge $U(1)$ group between branches with forward/backward evolutions on the Keldysh contour, as illustrated in Figure~\ref{fig:schemticas} (b). This leads to $F(\phi,Q_A)\sim \phi^2\log |A|$ when $G_{ud}\neq 0$, corresponding to both the volume-law entangled entropy phase for interacting systems and the critical phase for non-interacting systems. For $G_{ud}= 0$, $F(\phi,Q_A)$ satisfies an area-law as the entanglement entropy. A pictorial illustration is presented in Figure~\ref{fig:schemticas} (c), which clearly shows the FCS can be used to probe the entanglement phase transition of non-Hermitian Hamiltonians. We also remark on generalizations to systems with discrete symmetry, in higher dimensions, or with long-range hopping. 

\section{Model \& Setup}
We consider the non-Hermitian complex SYK chains with Brownian couplings. The total Hamiltonian $H=H_R-i H_I$ reads
     \begin{equation}\label{eq:H}
     \begin{aligned}
     H_R=&\sum_{ijx}\left[J_{ij}^x(t) c_{ix}^\dagger c_{jx+1}^{}+\text{h.c.}\right]+\sum_{ijklx}\frac{V_{ij,kl}^x(t)}{4}c_{ix}^\dagger c_{jx}^\dagger c_{kx}^{} c_{lx}^{},\\
     H_I=&\zeta\sum_{ix}(-1)^{x-1}c^\dagger_{ix}c^{}_{ix}.
     \end{aligned}
     \end{equation}
     Here $i\in\{1,2,...,N\}$ labels different fermion modes $c_{ix}$ on each site $x\in\{1,2,...,L\}$. The total charge $Q_c=\sum_{ix}c^\dagger_{ix}c^{}_{ix}$ is conserved under the evolution. $H_R$ contains random hopping $J_{ij}^x$ between nearest neighbor sites and random on-site interactions $V_{ij,kl}^x$, which are independent Brownian variables with 
     \begin{equation}
     \begin{aligned}
     \overline{J_{ij}^x(t_1)J_{ij}^x(t_2)}&=\frac{J\delta(t_{12})}{2N},\ \ \
\overline{V_{ij,kl}^x(t_1)V_{ij,kl}^x(t_2)}&=\frac{2V\delta(t_{12})}{N^3}.
     \end{aligned}
     \end{equation}
     $H_I$ describes a staggered imaginary potential with depth $\zeta$. It can be realized by performing weak measurements and following quantum trajectories without quantum jumps \cite{jianMeasurementInducedPhaseTransition2021a,zhangInformationScramblingEntanglement2023}. As an example, let us consider weak measurements for operator $O$, which is described by Kraus operators:
\begin{equation}
K^0_O=1-{\gamma_O} O^\dagger O+O(\gamma^2),\ \ \ \ \ K^1_O=\sqrt{2\gamma_O}O,
\end{equation}
where we have assumed $\gamma \ll 1$. We perform forced measurement by post-selection of outcome $0$. Introducing $\gamma_O=\zeta_O \delta t$, the evolution of $\rho$ due to the measurement then takes the form of imaginary-time evolutions
\begin{equation}
\rho(t+\delta t)\propto e^{-h_I\delta t}\rho(t)e^{-h_I\delta t},
\end{equation}
with $h_I=\zeta_O O^\dagger O$. Adding contributions from measurements with different $O$ and contributions from the unitary part, the total evolution is governed by the non-Hermitian Hamiltonian 
\begin{equation}
H=H_R-iH_I,\ \ \ \ \ \ H_I=\sum_O\zeta_O O^\dagger O.
\end{equation}
Choosing $O=c^\dagger_{ix}c^{}_{ix}$ or $c^{}_{ix}c^\dagger_{ix}$ for odd or even sites respectively with $\zeta_O=\zeta$, and following quantum trajectories without quantum jumps leads to the model introduced in Eq. \eqref{eq:H}.

     We are interested in computing the FCS on the steady state of the non-Hermitian dynamics. In this work, we prepare the system in a TFD state, as in the study of entanglement phase transitions in SYK-like models. 
     The definition of the TFD state requires introducing an auxiliary fermion system with annihilation operators $\eta_{ix}$. We first construct an EPR state between $c_{ix}$ and $\eta_{ix}$ in the occupation basis as $|\text{EPR}\rangle=\otimes_{ix}\frac{1}{\sqrt{2}}(|00\rangle_{ix}+|11\rangle_{ix})$. The TFD state is obtained after adding imaginary time evolutions to the EPR state $|\text{TFD}\rangle=\sqrt{Z^{-1}}e^{-\frac{\mu}{2}Q_c}|\text{EPR}\rangle$, where $Z$ is a normalization factor.

     For a realization of Brownian variables, the state at time $T$ is given by 
     \begin{equation}
     |\psi(T)\rangle= \frac{e^{-iH T}|\text{TFD}\rangle}{\sqrt{\langle\text{TFD}|e^{iH^\dagger T}e^{-iHT}|\text{TFD}\rangle}}.
     \end{equation}
     For the FCS, we choose a bipartition of the total system into $A$ and $\bar{A}$, where $A$ contains fermion modes $c_{ix}$ and $\eta_{ix}$ with $i\in\{1,2,...,|A|\}$. We further take $Q_A\equiv Q_{cA}-Q_{\eta A}=\sum_x c^\dagger_{ix} c^{}_{ix}-\eta^\dagger_{ix} \eta^{}_{ix}$, which annihilates the initial state as $Q_A|\text{TFD}\rangle=0$. The FCS \eqref{eq:defFCS} then reads
    \begin{equation}\label{eqn:FCSTFDdef}
    Z(\phi,Q_A)=\frac{\text{tr}_c[e^{iH^\dagger T}\mathcal{T}_ce^{-iHT}e^{-\mu Q_c/2}\mathcal{T}_c^\dagger e^{-\mu Q_c/2}]}{\text{tr}_c[e^{iH^\dagger T}e^{-iHT}e^{-\mu Q_c}]}.
    \end{equation}
    Here the trace is over the Hilbert space of fermion $c_{ix}$. The FCS takes the form of a correlator $\langle \mathcal{T}_c(T) \mathcal{T}_c^\dagger(0) \rangle_{\rho_c}$ with $\mathcal{T}_c=e^{i\phi Q_{cA}}$ on the ensemble of $\rho_c=Z^{-1}e^{-\mu Q_c}$. This is a close analog of the R\'enyi entropy calculation, as illustrated in Figure~\ref{fig:schemticas} (b). Moreover, for even $L$ the FCS is symmetric across $|A|=L/2$. This is due to the invariance of \eqref{eq:H} under a combination of the particle-hole transformation and the spatial reflection $c_{ix} \leftrightarrow c^\dagger_{iL-x}$, which gives $Z(\phi,Q_A)=Z(-\phi,Q_{\bar{A'}})=Z(\phi,Q_{A'})$ with $|A|=|\bar{A'}|=L-|A'|$.

    After computing the FCS for given random couplings, we need to perform the disorder average, which requires introducing disorder replicas. In SYK-like models, it is known that the saddle-point solution is replica-diagonal \cite{Kitaev:2017awl,Gu:2019jub}. Consequently, we can make the approximation $Z(\phi,Q_A)=\mathcal{Z}(\phi,Q_A)/\mathcal{Z}(0,Q_A)$, with $\mathcal{Z}(\phi,Q_A)=\overline{\text{tr}[e^{iH^\dagger T}\mathcal{T}_ce^{-iHT}e^{-\mu Q_c/2}\mathcal{T}_c^\dagger e^{-\mu Q_c/2}]}$. In the following subsections, we begin with an analysis of $\mathcal{Z}(0,Q_A)$, and then develop an effective theory for computing the response of twist operators for finite $\phi$.

     {There are two main reasons for selecting the TFD state. Firstly, measuring the relative charge $Q_{cA}-Q_{\eta A}$ in the doubled system, as shown in Figure~\ref{fig:schemticas} (b), is equivalent to measuring the charge $Q_{cA}$ at two different times. This allows us to directly interpret the FCS as the statistics of the charge transfer across the boundary $\partial A$. Secondly, the TFD state can be easily represented by a continuous boundary condition in the path-integral approach, which makes it simpler to numerically verify our results. However, we want to emphasize that our analysis yields qualitative features that should hold for more general initial states from a symmetry perspective.
     } 

\section{Saddle-point Solution}
In the large-$N$ limit, $\mathcal{Z}(\phi,Q_A)$ can be analyzed using the saddle-point approximation. We first focus on $\phi=0$. The Green's functions are defined as $G_x^{ab}(t,t')=\langle c_{ix}^a(t)\bar{c}_{ix}^b(t') \rangle$, where $a,b\in \{u,d\}$ labels fermion fields on branches with forward/backward evolutions. In SYK-like models, the path-integral of fermions can be transformed into a theory of collective fields $(G_x^{ab},\Sigma_x^{ab})$, in which the saddle-point equation is equivalent to the Schwinger-Dyson equation:
    \begin{equation}\label{eq:saddlepoint}
    \left(-if^a\partial_{t}\delta^{ac}+\zeta(-1)^{x-1}\delta^{ac}-\Sigma^{ac}_x\right)\circ G_{x}^{cb}=\delta^{ab}\hat I,
    \end{equation}
    where the summation over $c$ is implicit. We have $f^{u/d}=\pm i$ and self-energy 
    \begin{equation}
    \Sigma^{ac}_x=f^af^c \hat I\left[\frac{J}{2}(G^{ac}_{x+1}+G^{ac}_{x-1})-V(G^{ac}_{x})^2(G^{ca}_{x})^T\right].
    \end{equation}
    Here we have viewed both $G$ and $\Sigma$ as matrices in the time domain and defined the identity matrix $\hat I(t,t')=\delta(t-t')$. The transpose is applied in the time domain, and the contour indexes are explicitly shown. Away from the boundary of branches at $t=T$ and $t=0$, the Green's functions are time translational invariant $G_x^{ab}(t,t')=G_x^{ab}(t-t')$, which can be verified numerically. The solution can be obtained analytically, parametrized by $(\mathcal{P},\mathcal{S},z)$:
    \begin{equation}\label{eq:saddlesolution}
    \tilde{G}_x(\omega)=
    \begin{pmatrix}
    -i\omega+(-1)^{x-1} \mathcal{P}& -z^{-1}\mathcal{S}\\
     z \mathcal{S}&i\omega+(-1)^{x-1} \mathcal{P}
    \end{pmatrix}^{-1}.
    \end{equation}
    Here $\tilde{G}_x(\omega)=\int \frac{d\omega}{2\pi}e^{-i\omega t} G_x(t)$. $z=e^{\mu/2}$ is determined by the initial density matrix $\rho$. For $\zeta\geq J$, the solution is
    \begin{equation}\label{eq:saddlesolutionzetalarge}
    \mathcal{P}=\zeta-{J}/{2},\ \ \ \ \ \ \mathcal{S}=0,
    \end{equation}
    which gives a vanishing correlation between two branches $G^{ud}=0$. Following the analysis in previous studies, this corresponds to the area-law entangled phase. For $\zeta<J$, we instead have $\mathcal{P}=\zeta/{2}$ and
    \begin{equation}\label{eq:saddlesolutionzetasmall}
    1=\frac{J}{2}\frac{1}{\sqrt{\mathcal{P}^2+\mathcal{S}^2}}+\frac{V}{8}\frac{\mathcal{S}^2}{(\mathcal{P}^2+\mathcal{S}^2)^{3/2}}.
    \end{equation}
     Since $G_{ud}\neq 0$, this is a critical phase for $V=0$ and a volume-law entangled phase for $V>0$, as illustrated in Figure~\ref{fig:schemticas} (c). For $V=0$, we find $\mathcal{S}=\frac{J}{2}\sqrt{1-\zeta^2/J^2}$. For $V<2J$, $\mathcal{S}$ increases from 0 to $J/2+V/8$ continuously when we tune $\zeta$ from $J$ to $0$. For $V>2J$, the transition at $\zeta=J$ becomes the first order.
   \begin{figure}[tb]
        \centering
        \includegraphics[width=0.8\linewidth]{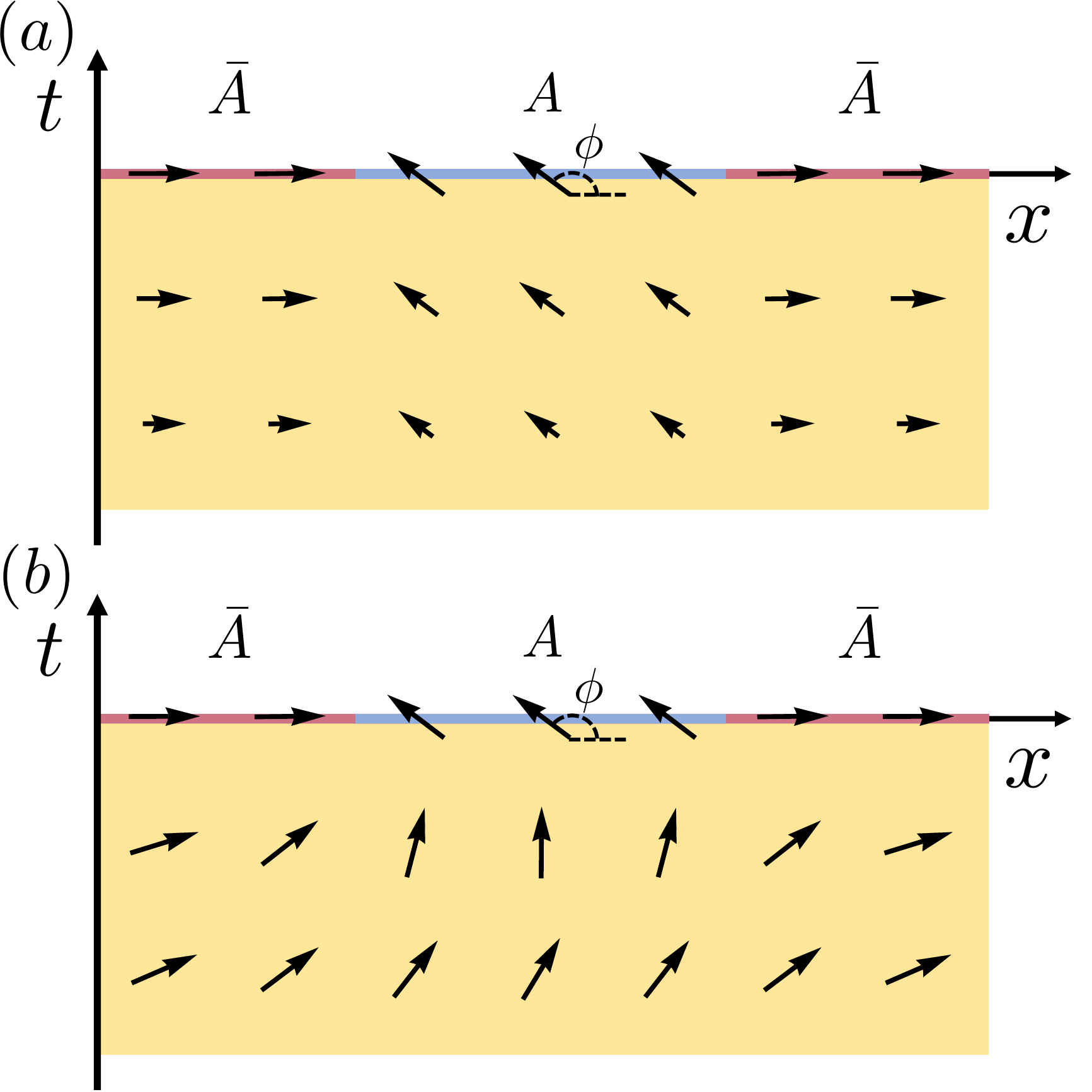}
        \caption{An illustration for the typical configurations of $G^{ud}_x(t,t)=|G^{ud}_x(t,t)|e^{-i\varphi(x,t)}$ in calculation of the FCS. Since $\varphi(x,t)$ is defined as a phase field, we represent it as the angle of a 2D vector (spin). Furthermore, the magnitudes of spins represent $|G^{ud}_x(t,t)|$. (a). For $\zeta>J$, the $F(\phi,Q_A)$ can be estimated by a perturbative calculation of the hopping term $J$, which gives $F(\phi,Q_A)\sim (1-\cos \phi) |\partial A|$. (b). For $\zeta<J$, the twist operators create a $\phi$-vortex pair, whose the excitation energy gives $F(\phi,Q_A)\sim \phi^2 \log |A|$. There is also a similar contribution near $t=0$. }
        \label{fig:spin}
    \end{figure}

\section{Relative Phase Twist}
We ask how the insertion of twist operators $\mathcal{T}_r$ changes the saddle-point solution. We begin with the simplest case where $\bar{A}=\emptyset$ and $Q_{cA}=Q_c$ is the conserved charge. The twist operator then commutes with the Hamiltonian $H$. As a result, it only induces a relative phase rotation between $u$ and $d$ branches $(c^u_{ix},c^d_{ix})\rightarrow (c^u_{ix}e^{-i\phi},c^d_{ix})$, which gives 
    \begin{equation}\label{eqn:LA=L}
    \begin{aligned}
    G^{du}_x(t,t')_\phi&=\langle c_{ix}^d(t)\mathcal{T}_c(T)\bar{c}_{ix}^u(t') \mathcal{T}_c^\dagger(0) \rangle_\rho=e^{i\phi}G^{du}_x(t,t').
    \end{aligned}
    \end{equation}
    Here $G^{du}_x(t,t')_\phi$ is the Green's function with twist operators. Similarly, we have $G^{ud}_x(t,t')_\phi=e^{-i\phi} G^{ud}_x(t,t')$. This reveals that the twist operator is coupled to the phase fluctuation in the off-diagonal components of Green's functions. 

We then consider a general subsystem size $|A|$. Due to the presence of twist operators, the system no longer exhibits translation symmetry and an exact solution of the saddle-point equation is unavailable. However, for large $L>|A|\gg1$, we expect only soft modes can be excited. For $G^{ud}= 0$, there is no symmetry reason for the existence of any soft mode, and the correlation in the system is generally short-ranged. As a result, we expect $F(\phi,Q_A)$ to satisfy an area law. For $G^{ud}\neq 0$ the soft mode in the system is just the relative phase mode: The saddle-point equation \eqref{eq:saddlepoint} is invariant under the relative phase rotation, while the solution \eqref{eq:saddlesolution} breaks the symmetry when $G^{ud}\neq 0$. As a result, the relative phase rotation becomes a Goldstone mode. This indicates we can approximate 
    \begin{equation}
    G_x(t,t)_\phi\approx  \begin{pmatrix}
    G_x^{uu}(0)&e^{-i\varphi(x,t)}G^{ud}(0)\\
    e^{i\varphi(x,t)}G^{du}(0)&G_x^{dd}(0)
    \end{pmatrix}.  
    \end{equation}
    The FCS is then determined by minimizing the effective action of $\varphi(x,t)$ with the boundary condition specified by the twist operator:
    \begin{equation}\label{eq:bdycond}
    \varphi(x,T)=\begin{cases}
    0 & x\in \bar{A},\\
    \phi & x\in A.
    \end{cases}
    \end{equation}
    This can be derived by noticing $G_x^{du}(T,T)_\phi=G_x^{uu}(T,T)_\phi$ for $x\in \bar{A}$ and $G_x^{du}(T,T)_\phi=e^{i\phi}G_x^{uu}(T,T)_\phi$ for $x\in A$. 
\section{Effective Action \& FCS}

In this section, we explicitly derive the effective action governing the fluctuation around the saddle-point solution, which justifies our analysis above and gives closed-form expressions for $F(\phi,Q_A)$.
    
    We first consider the short-range entangled phase with $\zeta>J$. The effective action is given by expanding the $G$-$\Sigma$ action around the saddle-point \eqref{eq:saddlesolution} and \eqref{eq:saddlesolutionzetalarge} \cite{maldacena2016remarks}. Leaving details into the Appendix \ref{app:der}, to the quadratic order we find
    \begin{equation}\label{eqn:S1}
   \frac{S_\text{eff}}{N} =  \int_{\Omega, k}
        \mathfrak{g}^{du}_{-k}
        \left(
        \begin{array}{cc}
            -2 \zeta +2J-\frac{Jk^2}{2} & {i\Omega}   \\
             {i \Omega}  & -2 \zeta 
        \end{array}
        \right)
        \mathfrak{g}^{ud}_{k},
    \end{equation}
    where we have introduced $\mathfrak{g}_k^{ud} = \left( \delta G^{ud}_{k}, \delta G^{ud}_{k+\pi} \right)^T$ and  $\mathfrak{g}_k^{du} = \left( \delta G^{du}_{k}, \delta G^{du}_{k+\pi} \right)$. To estimate $F(\phi,Q_A)$, we take a perturbation approach in terms of small $J$. Although $G^{ud}=0$ for $0\ll t\ll T$, it becomes finite near the boundaries due to the boundary condition of the contour. Without the hopping term $J$, different sites decouple and $G^{ud}$ can be computed as in \eqref{eqn:LA=L}, which gives $\varphi(x,t)=\phi$ for $x\in A$ and $\varphi(x,t)=0$ for $x\in \bar{A}$. This is illustrated in Figure~\ref{fig:spin} (a). To further estimate the decay of $|G^{ud}|$, we integrate out $\mathfrak{g}_k^{du}$ in \eqref{eqn:S1}. For small $J$, this imposes the constraint
    \begin{equation}
     \left(-\frac{1}{2\zeta}\partial_t^2+2\zeta-2J\right) \delta G^{ud}_x(t,t)=0
    \end{equation}  
    which gives $|G^{ud}_x(t,t)|\sim e^{-2(T-t)\sqrt{\zeta(\zeta-J)}}$. Similar calculation works for $|G^{du}_x(t,t)|$. Then we can compute contributions from small hopping terms in $\mathcal{Z}(\phi,Q_A)$ as
    \begin{equation}
    {J}N~\text{Re}\int dt~ G_{A}^{ud}G_{\bar{A}}^{du}\sim -\frac{J}{\sqrt{{\zeta(\zeta-J)}}}N\cos\phi.
    \end{equation} 
    Subtracting the contribution from $\mathcal{Z}(0,Q_A)$, we finally obtain 
    \begin{equation} \label{eq:Farea}
    F(\phi,Q_A)\sim\frac{J}{\sqrt{{\zeta(\zeta-J)}}}N(1-\cos\phi).
    \end{equation}

    Now we consider the long-range correlated phase with $\zeta<J$. As explained in the last section, we need to derive the effective theory for the relative phase mode $\varphi(x,t)$. Similar to the derivation of \eqref{eqn:S1}, we now expand the $G$-$\Sigma$ action around the saddle-point \eqref{eq:saddlesolution} and \eqref{eq:saddlesolutionzetasmall} to the quadratic order, with the identification that $\delta G^{ud}_x(t,t)= -i\varphi(x,t)G^{ud}(0)$ and $\delta G^{du}_x(t,t)= i\varphi(x,t)G^{du}(0)$. As derived in the Appendix \ref{app:der}, the result reads 
    \begin{equation}\label{eq:action2_main}
        \begin{split}
           S_{\text{eff}}&=\frac{N\mathcal{S}^2}{4(\zeta ^2+4 \mathcal{S}^2)}\int_{x,t}~\Bigg(\frac{(\partial_t\varphi)^2 }{2 \sqrt{\zeta ^2+4 \mathcal{S}^2}-J} +J(\partial_x\varphi)^2\Bigg). \\
        \end{split}
    \end{equation}
    This is a large-$N$ XY model. Consequently, the system exhibits an emergent conformal symmetry. Unlike the emergent replica conformal symmetry in non-Hermitian free-fermion systems, here the conformal symmetry is a consequence of charge $U(1)$ symmetry, which is stable against adding interactions. The boundary condition in \eqref{eq:bdycond} then excites a $\phi$-vortex pair, as sketched in Figure~\ref{fig:spin} (b). The FCS is equal to the excitation of the vortex pair. In the limit of $L\rightarrow \infty$, the result reads \cite{zhai_2021,zhangUniversalEntanglementTransitions2022}
    \begin{equation} \label{eq:Flong}
    F(\phi,Q_A)\sim  \frac{\mathcal{S}^2\phi^2N}{\zeta ^2+4 \mathcal{S}^2}\left(\frac{J}{2 \sqrt{\zeta ^2+4 \mathcal{S}^2}-J}\right)^{1/2}\log |A|.
    \end{equation}
    In particular, it shows non-analyticity near $\phi=\pi$. For finite $L$, $\log|A|$ should be replaced by $\log [L\sin(\pi|A|/L)/\pi]$ due to the conformal invariance. Comparing \eqref{eq:Farea} and \eqref{eq:Flong}, we find $F(\phi,Q_A)$ shows qualitative different scalings of both $|A|$ and $\phi$ for both interacting and non-interacting systems, and thus serves as a universal probe of the entanglement phase transition in non-Hermitian Hamiltonian dynamics.

   \begin{figure}[tb]
        \centering
        \includegraphics[width=0.98\linewidth]{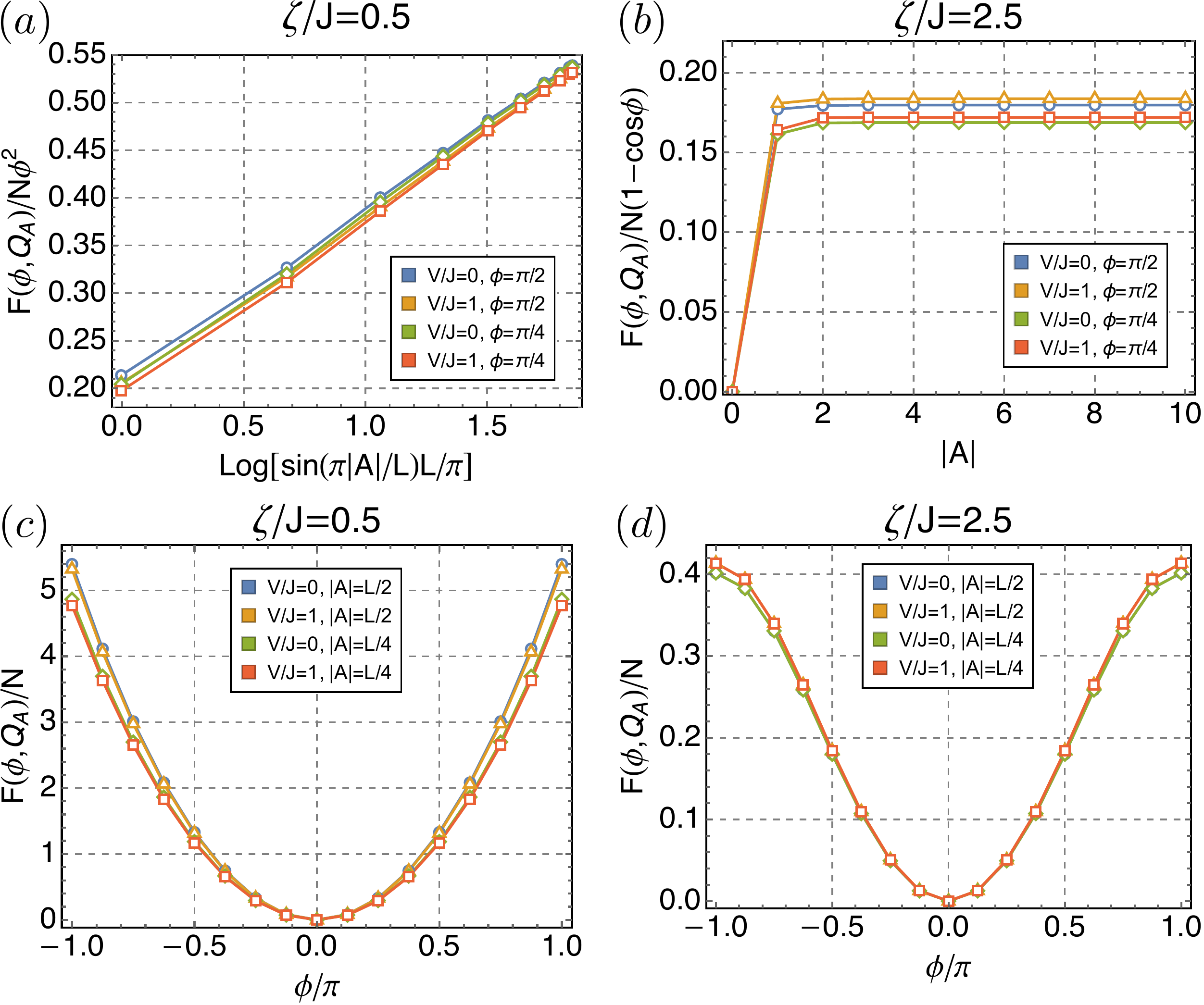}
        \caption{The numerical calculation of the FCS in the large-$N$ limit by solving the Schwinger-Dyson equation with $\mu=0.5$. (a-b). $F(\phi,Q_A)$ for different subsystem size $|A|$. The results show $F(\phi,Q_A) \propto \log|A|$ for $\zeta/J=0.5$ and $F(\phi,Q_A)\propto |\partial A|$ for $\zeta/J=2.5$. (c-d). $F(\phi,Q_A)$ for different twist strength $\phi$. The results show $F(\phi,Q_A) \propto \phi^2$ for $\zeta/J=0.5$ and $F(\phi,Q_A)\propto (1-\cos\phi)$ for $\zeta/J=2.5$. }
        \label{fig:num}
    \end{figure}

To justify our theoretical predictions \eqref{eq:Farea} and \eqref{eq:Flong}, we numerically study the FCS in the large-$N$ limit by solving the saddle-point equation with twist operators $\mathcal{T}_c$ and computing the on-shell action. Similar approaches have been widely adopted to simulate dynamics of R\'enyi entropies in SYK-like models \cite{liuNonunitaryDynamicsSachdevYeKitaev2021,zhang2021emergent,zhang2020entanglement,zhang2022quantum}. In Figure~\ref{fig:num}, we present results for $L=20$ with $\mu=0.5$. For $\zeta<J$, we check that $F(\phi,Q_A)$ is a linear function of $\log [L\sin(\pi|A|/L)/\pi]$ and is a quadratic function of $\phi$. Near $\phi=\pm \pi$, we have two different saddle point solutions, which leads to a non-analyticity. For $\zeta>J$, we check that $F(\phi,Q_A)$ shows area-law behavior for large $|A|$, and is proportional to $(1-\cos \phi)$. All results are tested for both the interacting case ($V=J$) and the non-interacting case ($V=0$).

\section{Discussions}

In this work, we study the full counting statistics on the steady state of non-Hermitian complex SYK models. We find $F(\phi,Q_A)$ show different scaling of both $|A|$ and $\phi$ in phases with different entanglement properties. Using an effective spin model, we show that: For $\zeta>J$, the system is short-range correlated. We can approximate $F(\phi,Q_A)\sim (1-\cos \phi)|\partial A|$ for $\zeta \gg J$. For $\zeta<J$, the system is long-range correlated with $F(\phi,Q_A)\sim \phi^2\log |A|$, which exhibits non-analyticity near $\phi=\pi$. We further validate our theoretical predictions by numerically solving the saddle-point equation. 
    
    We point out the identification of the relative phase twist is not restricted to the TFD states, but valid for general initial states that are eigenstates of local charge operators. Without loss of generality, we assume $Q_A|\psi\rangle=0$. This gives $Z(\phi,Q_A)\propto \langle\psi|  e^{iH^\dagger T}e^{i\phi Q_A}e^{-iHT}|\psi\rangle =\langle\psi|  e^{iH^\dagger T}e^{i\phi Q_A}e^{-iHT}e^{-i\phi Q_A}|\psi\rangle $. Similar to \eqref{eqn:FCSTFDdef}, a pair of twist operators appear, which leads to a relative phase twist between forward evolution and backward evolution branches. Moreover, there is no difference between the TFD state and more general initial states from the symmetry perspective{: In the absence of boundary conditions at $t=0$ and $t=T$, fermion fields on distinct branches of the system can undergo independent transformations. As a result, the system exhibits a $U(1)\otimes U(1)$ symmetry. However, in the volume law phase where $G_{ud}\neq 0$, this symmetry is broken down to a single $U(1)$ symmetry, leading to the emergence of a Goldstone mode.} As a result, we believe our theoretical predictions reveal universal features of the FCS across the entanglement transition of non-Hermitian Hamiltonians.

    Several additional remarks are in order: Firstly, instead of systems with $U(1)$ symmetry, we can consider models with discrete symmetries. It is then natural to expect the phase with small non-Hermitian strength is still symmetry breaking phase, but with no Goldstone mode. The dominant contribution now becomes a domain wall, instead of a vortex pair. This is consistent with recent numerics in \cite{tirritoFullCountingStatistics2023}. Secondly, our results can be generalized to higher dimensions straightforwardly. As an example, let us consider a non-Hermitian complex SYK chain in 2D. The short-range correlated phase with $\zeta>J$ is still area law entangled with $F(\phi,Q_A)\sim (1-\cos \phi)R$ for a subsystem $A$ with radius $R$. For the phase with $\zeta<J$, the effective theory becomes a 3D XY model. The vortex pair is then replaced by a vortex ring. Consequently, we have $F(\phi,Q_A)\sim \phi^2 R\log R$. Finally, in experimental systems, long-range interactions may present. If we consider a long-range hopping term that decays as $1/r^{\alpha}$ as in \cite{zhangUniversalEntanglementTransitions2022}, a fractal phase with $F(\phi,Q_A)\sim \phi^2 L_A^{1-z}$ appears with $z=\frac{2\alpha-1}{2}$ for $\alpha\in (0.5,1.5)$. 

\emph{Acknowledgement.} 
We thank Hui Zhai for bringing our attention to the full counting statistics. We thank Meng Cheng, Yingfei Gu, and Shang-Qiang Ning for helpful discussions. PZ would like to thank Xiao Chen, Shao-Kai Jian, and Chunxiao Liu for valuable discussions during the collaboration on related topics. The project is supported by NSFC under Grant No. 12374477.

\appendix

\onecolumngrid
\section{The derivation of the full counting statistics on the TFD state}\label{app:TFD}
The FCS is defined as 
    \begin{equation}
    \mathcal{Z}(\phi,Q_A)\equiv\text{tr}_{c,\eta}[\rho(T)e^{-i\phi Q_A}]
    \end{equation}
    with $Q_A\equiv Q_{cA}-Q_{\eta A}=\sum_x c^\dagger_{ix} c^{}_{ix}-\eta^\dagger_{ix} \eta^{}_{ix}$.
    Here, the trace is over the $c$ fermion system and $\eta$ fermion system, and  $\rho(T)=\frac{1}{Z(T)}e^{-iHT}\rho(0)e^{iH^\dagger T}$ is the density matrix at time $T$. The $Z(T)=\text{tr}[e^{-iHT}e^{-i\mu Q_c}e^{iH^\dagger T}]$ is  the normalization factor. The initial state is the $|\text{TFD}\rangle=\sqrt{Z^{-1}}e^{-\frac{\mu}{2}Q_c}|\text{EPR}\rangle$. Put all the definitions into the FCS, we obtain
    \begin{equation}
       \begin{split}
        \mathcal{Z}(\phi,Q_A)=&\frac{1}{Z(T)}\text{tr}_{c,\eta}[e^{-iHT}|\text{TFD}\rangle \langle\text{TFD}| e^{iH^\dagger T}e^{-i\phi Q_A}]\\
        =& \frac{1}{Z(T)}\langle\text{TFD}|e^{iH^\dagger T}e^{-i\phi Q_A}e^{-iHT}|\text{TFD}\rangle \\
        =&\frac{1}{Z(T)}\langle\text{TFD}|e^{iH^\dagger T}e^{-i\phi Q_{cA}}e^{i\phi Q_{\eta A}}e^{-iHT}|\text{TFD}\rangle.
        \end{split}
    \end{equation}
    Since the Hamiltonian $H$ is defined on the $c$ fermion system and $Q_{\eta A}$ is defined on the auxiliary $\eta$ fermion system, the operator $H$ and $Q_{\eta A}$ commute with each other. Thus, we have
    \begin{equation}
        \begin{split}
        \mathcal{Z}(\phi,Q_A)
        =&\frac{1}{Z(T)}\langle\text{TFD}|e^{iH^\dagger T}e^{-i\phi Q_{cA}}e^{-iHT}e^{i\phi Q_{\eta A}}|\text{TFD}\rangle.
       \end{split}
    \end{equation}
    Given that $Q_A$ annihilates the initial TFD state $Q_A|\text{TFD}\rangle=0$, we obtain $e^{i\phi Q_A}|\text{TFD}\rangle=|\text{TFD}\rangle$. Thus, we further have $e^{i\phi Q_{cA}}|\text{TFD}\rangle=e^{i\phi Q_{\eta A}}|\text{TFD}\rangle$. Therefore, FCS can be further written as 
    \begin{equation}
        \begin{split}
        \mathcal{Z}(\phi,Q_A)
        =&\frac{1}{Z(T)}\langle\text{TFD}|e^{iH^\dagger T}e^{-i\phi Q_{cA}}e^{-iHT}e^{i\phi Q_{c A}}|\text{TFD}\rangle\\
        =&\frac{1}{Z(T)}\langle\text{TFD}|\mathcal{T}_c(T)\mathcal{T}^\dagger_c(0)|\text{TFD}\rangle
       \end{split}
    \end{equation}
    with $\mathcal{T}_c=e^{-i\phi Q_{cA}}$. Therefore, we find that the FCS takes the form of a correlator. We also have
    \begin{equation}
       \begin{split}
        \mathcal{Z}(\phi,Q_A)
        =&\frac{1}{Z(T)}\langle\text{TFD}|e^{iH^\dagger T}e^{-i\phi Q_{cA}}e^{-iHT}e^{i\phi Q_{c A}}|\text{TFD}\rangle\\
        =&\frac{1}{Z(T)}\langle\text{EPR}|e^{-\frac{\mu}{2} Q_c}e^{iH^\dagger T}e^{-i\phi Q_{cA}}e^{-iHT}e^{i\phi Q_{c A}}e^{-\frac{\mu}{2} Q_c}|\text{EPR}\rangle\\
        =&\frac{1}{Z(T)}\text{tr}_c[e^{-\frac{\mu}{2} Q_c}e^{iH^\dagger T}e^{-i\phi Q_{cA}}e^{-iHT}e^{i\phi Q_{c A}}e^{-\frac{\mu}{2} Q_c}]\\
        =&\frac{1}{Z(T)}\text{tr}_c[e^{iH^\dagger T}e^{-i\phi Q_{cA}}e^{-iHT}e^{i\phi Q_{c A}}e^{-\mu Q_c}]\\
        =&\frac{1}{Z(T)}\text{tr}_c[e^{iH^\dagger T}\mathcal{T}_c e^{-iHT}\mathcal{T}_c^\dagger e^{-\mu Q_c}]\\
        =&\frac{1}{Z(T)}\text{tr}_c[e^{iH^\dagger T}\mathcal{T}_c e^{-iHT} e^{-\frac{\mu}{2}Q_c}\mathcal{T}_c^\dagger e^{-\frac{\mu}{2}Q_c}].
        \end{split}
    \end{equation}
    We have used the $\mathcal{T}_c^\dagger$ and $Q_c$ commute with each other to obtain the last equality. This is the Eq.(5) in the main text. Notice that here the trace is over the $c$ fermion system. 
     
     \section{The path-integral representation of the full counting statistics}\label{app:action}
     As explained in the main text, we focus on the path-integral representation of the full counting statistics (FCS). The time evolution of a FCS $\mathcal{Z}(\phi,Q_A)$ is
     \begin{equation}\label{eq:Z_phi_QA}
     \mathcal{Z}(\phi,Q_A)=\overline{\text{tr}[e^{iH^\dagger T}\mathcal{T}_ce^{-iHT}e^{-\mu Q_c/2}\mathcal{T}_c^\dagger e^{-\mu Q_c/2}]}
     \end{equation}
     and it does not preserve normalization. To evaluate $\mathcal{Z}(\phi,Q_A)$ using path integral, one needs two contours similar to the Keldysh contour.
     Denoted by $u$ and $d$ for forward and backward evolution, the action on these two contours schematically is
     \begin{eqnarray}
     -I = \int dt \sum_{x,a} \left( -\iu f^a \bar{c}_x^a  \partial_t c_x^a + f^a \left( J^{x}(t) \bar{c}_x^a c_{x+1}^a + J^{x}(t) \bar{c}_{x+1}^a c_{x}^a \right) + f^a \frac{V^x(t)}{4} \bar{c}_x^a \bar{c}_x^a c_x^a c_x^a +
    \zeta (-1)^{x-1} \bar{c}_{x}^a c_{x}^a
      \right), \nonumber \\ 
     \end{eqnarray}
     where $a = u, d$ denotes two contours, and the superscript for fermion species on each site is suppressed. Here $f^{u/d}=\pm i$. The last term in the equation doesn't have a contour-dependent prefactor, because of the non-unitary $H_I$ and the definition of Eq.~\eqref{eq:Z_phi_QA}.
     The effective action on these two contours after integrating out disorder and introducing $1=\int \diff t \diff t' \Sigma^{ab}_x(t, t') (G^{ba}_x(t',t) - c_x^b(t')  \bar{c}_x^a(t) )$ is
     \begin{eqnarray} \label{eq:full_action}
     - \frac{I}{N} =&& \sum_x  \Tr \log \Big( -\iu f^a \partial_t \delta^{ac} + \zeta (-1)^{x-1}\delta^{ac} - \Sigma_x^{ac} \Big)  + \int dtdt'~\Big[ \Sigma_x^{ac} (G_x^{ca})^T \nonumber \\
     &&+ f^a f^c \hat{I} \left(  \frac{V}{4} (G_x^{ac})^2 \left((-G_x^{ca})^2\right)^T + \frac{J}{4} \left(G^{ac}_{x+1} (-G^{ca}_{x})^T + G^{ac}_{x-1} (-G^{ca}_{x})^T\right)\right) \Big],
     \end{eqnarray}
     where $G^{ab}$ and $\Sigma^{ab}$ are the bilocal fields with time arguments $t$ and $t'$ omitted, which characterize the two-point function of Majorana fermions or the corresponding self-energy at $a$ and $b$ contours. Here $\hat{I}=\delta(t-t')$ means the Brownian condition.
     As a result, the saddle point equation is 
     \begin{equation}\label{eq:saddlepoint}
        \begin{split}
            &   \sum_c\left(-if^a\partial_{t}\delta^{ac}+\zeta(-1)^{x-1}\delta^{ac}-\Sigma^{ac}_x\right)\circ G_{x}^{cb}=\delta^{ab}\hat I, \\
            &   \Sigma^{ac}_x=f^af^c \hat I\left[V(G^{ac}_{x})^2(-G^{ca}_{x})^T+J(G^{ac}_{x+1}+G^{ac}_{x-1})/2\right]. \\
        \end{split}
     \end{equation}

     \section{The derivation of the effective action}\label{app:der}
     The effective action is given by expanding the $G-\Sigma$ action \eqref{eq:full_action} around the saddle-point solution. In this section, we give a detailed derivation for effective actions. 
     
     The solution can be obtained analytically, parametrized by $(\mathcal{P},\mathcal{S},z)$:
     \begin{equation}\label{eq:saddlesolution}
        \tilde{G}_x(\omega)=
        \begin{pmatrix}
            -i\omega+(-1)^{x-1} \mathcal{P}& -z^{-1}\mathcal{S}\\
            z \mathcal{S}&i\omega+(-1)^{x-1} \mathcal{P}
        \end{pmatrix}^{-1}.
     \end{equation}
     Here $\tilde{G}_x(\omega)=\int \frac{d\omega}{2\pi}e^{-i\omega t} G_x(t)$. $z=e^{\mu/2}$ is determined by the initial density matrix $\rho$. For $\zeta\geq J$, the solution is
     \begin{equation}\label{eq:saddlesolutionzetalarge}
        \mathcal{P}=\zeta-{J}/{2},\ \ \ \ \ \ \mathcal{S}=0,
     \end{equation}
     which gives a vanishing correlation between two branches $G^{ud}=0$. Following the analysis in previous studies, this corresponds to the area-law entangled phase. For $\zeta<J$, we instead have 
     \begin{equation}\label{eq:saddlesolutionzetasmall}
        \mathcal{P}=\zeta/{2},\ \ \ \ \ \ 1=\frac{J}{2}\frac{1}{\sqrt{\mathcal{P}^2+\mathcal{S}^2}}+\frac{V}{8}\frac{\mathcal{S}^2}{(\mathcal{P}^2+\mathcal{S}^2)^{3/2}}.
     \end{equation}
 
    Now we consider saddle-point fluctuations,
    \begin{eqnarray}\label{eq:fluctuation}
        \Sigma(t_1, t_2) = \Sigma_s(t_1, t_2) + \delta \Sigma(t_1) \delta(t_{12}), \quad G(t_1, t_2) = G_s(t_1, t_2) + \delta G(t_1, t_2).
    \end{eqnarray}
    We will separately evaluate the fluctuation in Eq.~\eqref{eq:full_action}, including $\Tr \log$ term, linear coupling $\Sigma G$ term, and interaction term related to $J$ and $V$. Now we discuss both $\zeta > J$ and $\zeta < J$ cases. 
    
    \subsection{$\zeta > J$}
    
    \paragraph{$\Tr \log term$}
    
    Expanding the $\Tr \log$ term leads to the following equation
    \begin{equation}\label{eq:tracelog}
        \begin{split}
            &\Tr \log \Big( -\iu f^a \partial_t \delta^{ac} + \zeta (-1)^{x-1}\delta^{ac} - \Sigma_x^{ac} \Big) \\
            =& \Tr \log \Big( (G^{-1}_s)_x^{ac} \Big) + \Tr \log \Big( \hat{I}^{ab} - (G_s)^{ca}_x \circ (\delta\Sigma_s)_x^{ab} \Big)  \\
            =& \Tr \log \Big( (G^{-1}_s)_x^{ac} \Big) - \int \diff \tau_1\diff\tau_2 \Big( (G_s)^{ca}_x(\tau_1,\tau_2)  (\delta\Sigma_s)_x^{ac}(\tau_2) \delta(\tau_2-\tau_1) \Big)  \\
            &- \int \diff \tau_1\diff\tau_2\diff\tau_3\diff\tau_4 \frac{1}{2} \Big( (G_s)^{ab}_x(\tau_1,\tau_2)  (\delta\Sigma_s)_x^{bc}(\tau_2) \delta(\tau_2-\tau_3) (G_s)^{cd}_x(\tau_3,\tau_4)  (\delta\Sigma_s)_x^{da}(\tau_4) \delta(\tau_4-\tau_1) \Big)  \\
        \end{split}
    \end{equation}
    The contour index $a,b,c,d$ implicitly used the Einstein summation. The second-order term can be transformed 
    \begin{equation}\label{eq:2ndorder}
        \begin{split}
            &- \int \diff \tau_1\diff\tau_2\diff\tau_3\diff\tau_4 \frac{1}{2} \Big( (G_s)^{ab}_x(\tau_1,\tau_2)  (\delta\Sigma_s)_x^{bc}(\tau_2) \delta(\tau_2-\tau_3) (G_s)^{cd}_x(\tau_3,\tau_4)  (\delta\Sigma_s)_x^{da}(\tau_4) \delta(\tau_4-\tau_1) \Big)  \\
            =&- \int \diff \tau_1\diff\tau_2 \frac{1}{2} \Big( (G_s)^{ab}_x(\tau_1,\tau_2)  (\delta\Sigma_s)_x^{bc}(\tau_2) (G_s)^{cd}_x(\tau_2,\tau_1)  (\delta\Sigma_s)_x^{da}(\tau_1)  \Big) \\
            =& - \frac12 \int_{\omega, \Omega} \Tr\left[ G_s(\omega + \Omega) \delta \Sigma(\Omega) G_s(\omega) \delta \Sigma(-\Omega) \right],\\
        \end{split}
    \end{equation}
    where $\int_\Omega \equiv \int^{\infty}_{-\infty} \frac{d\Omega}{2\pi} $, and we have used the Fourier transform $\delta \Sigma(\Omega) = \int dt \delta \Sigma(t) e^{i \Omega t}$. In the last step of Eq.~\eqref{eq:2ndorder}, $\int_\omega$ can be integrated out using residue theorem, and $G_s$ insert Eq.~\eqref{eq:saddlesolution}, \eqref{eq:saddlesolutionzetalarge}. Finally, we arrive at
    \begin{equation}\label{eq:M1}
            -\delta I_1^{(1)}/N = \sum_x \int_{\Omega} \sigma_x^{T}(\Omega) \mathcal{M}_1^{(1)} \sigma_x(-\Omega)\\
    \end{equation} for the trace log term.
    There are two independent diagonal fields $\delta \Sigma^{uu}, \delta \Sigma^{dd}$ and two independent off-diagonal fields $\delta \Sigma^{ud}, \delta \Sigma^{du}$. The full kernel implies that (a) off-diagonal fields decouple from the diagonal field and (b) two independent off-diagonal fields have nontrivial interactions.
    We denote these two nontrivial off-diagonal fields to be $\sigma_x = \left( \delta \Sigma^{ud}_x, \delta \Sigma^{du}_x \right)^T$, and the corresponding kernel reads
    \begin{eqnarray}
        \mathcal{M}_1^{(1)} = \frac{1}{2(\Omega^2 + (J-2\zeta)^2)} \left( \begin{array}{cc} 
            0 & J - 2 \zeta + \iu \Omega (-1)^x \\
            J - 2 \zeta - \iu \Omega (-1)^x & 0 \\
             \end{array} \right). 
    \end{eqnarray}
    
    \paragraph{$\Sigma G$ term}
    For the $\Sigma G$ term, 
    \begin{equation}\label{eq:SigmaG_1}
        \begin{split}
            &\int dtdt'~ \Sigma_x^{ac} (G_x^{ca})^T  \\
            =& -I_2^{(1)}/N +  \text{Diag. term} + \sum_x \frac{1}{2} \int_{\Omega} \left[ g_x^T(\Omega)  \begin{pmatrix}
                0 & 1 \\
                1 & 0 \\
            \end{pmatrix} \sigma_x(-\Omega) + \sigma^T_x(\Omega) 
        \begin{pmatrix}
            0 & 1 \\
            1 & 0 \\
        \end{pmatrix} g_x(-\Omega) \right] \\
        \equiv&-I_2^{(1)}/N - \delta I_2^{(1)}/N,
        \end{split}
    \end{equation}
     where $g_x = \left( \delta G^{ud}_x, \delta G^{du}_x \right)^T$ and $\sigma$ has been defined before. 
     
     \paragraph{$V$ term}
     \begin{equation}\label{eq:V_action1}
        \begin{split}
            &\int \diff t \diff t' \sum_x
            f^a f^c \hat{I}   \frac{V}{4} (G_x^{ac})^2 \left((-G_x^{ca})^2\right)^T \\
            =&  \text{Diag. term} + \frac{1}{4} \sum_x \int \diff t \left[ \left(G_x^{12}(t,t)\right)^2 \left(-G_x^{21}(t,t)\right)^2 
            + \left(G_x^{21}(t,t)\right)^2 \left(-G_x^{12}(t,t)\right)^2
            \right]. \\
        \end{split}
     \end{equation}
    After applying saddle point fluctuation Eq.~\eqref{eq:fluctuation} and keeping to second order of $\delta G$, we find all fluctuation terms are proportional to $G_s^{12}(\Omega)$ or $G_s^{21}(\Omega)$. According to condition Eq.~\eqref{eq:saddlesolutionzetalarge}, it means $V$ term doesn't contribute to action in the second order.
   
   \paragraph{$J$ term}
   \begin{equation}\label{eq:J_action1}
        \begin{split}
            &\int \diff t \diff t' \sum_x
            \frac{J}{4} \left[G^{ac}_{x+1} (-G^{ca}_{x})^T + G^{ac}_{x-1} (-G^{ca}_{x})^T\right] \\
            =& - I_4^{(1)}/N +  \text{Diag. term} - \int_{k} \int \diff t 
            \frac{J}{2} \left[ \delta G^{12}_{k}(t,t) \delta G^{21}_{-k}(t,t) \cos(k) + \delta G^{21}_{k}(t,t) \delta G^{12}_{-k}(t,t) \cos(k)\right] \\
            =& - I_4^{(1)}/N + \text{Diag. term} + \int_{k} \int \diff t 
            \frac{-J}{2} g_k^T(\Omega) 
            \begin{pmatrix}
                0 & \cos{k} \\
                \cos{k} & 0 \\
            \end{pmatrix}
             g_k(-\Omega)\\
             \equiv& - I_4^{(1)}/N - \delta I_4^{(1)}/N, \\
        \end{split}
   \end{equation}
    In the last step, we apply the saddle fluctuation solution and perform the Fourier transition on $k$, where $\delta G_x^{ab}(t,t) = \int_{k} G_x^{ab}(t,t) e^{\iu k x}$. Here $\int_{k} \equiv \int \frac{\diff k }{2\pi}$. 
    
    \paragraph{Effective Action}
    Sum all contributions Eq.~\eqref{eq:M1},~\eqref{eq:SigmaG_1},~\eqref{eq:J_action1} together and keep the second order off-diagonal terms, we arrive
    \begin{equation}\label{eq:deltaAction1_step1}
    \begin{aligned}
        - \delta I/N = \int_{\Omega} \Bigg\{ &
        \sum_x\left[  \sigma^{T}(\Omega) \mathcal{M}_1 \sigma(-\Omega) + \frac{1}{2}  \left[ g_x^T(\Omega)  \begin{pmatrix}
            0 & 1 \\
            1 & 0 \\
        \end{pmatrix} \sigma_x(-\Omega) + \sigma^T_x(\Omega) 
        \begin{pmatrix}
            0 & 1 \\
            1 & 0 \\
        \end{pmatrix} g_x(-\Omega) \right] \right] \\&+ \sum_k \frac{-J}{2} g_k^T(\Omega) 
        \begin{pmatrix}
            0 & \cos{k} \\
            \cos{k} & 0 \\
        \end{pmatrix}
        g_k(-\Omega)
         \Bigg\}.
         \end{aligned}
    \end{equation}
    Integrate out $\sigma$ leads to
    \begin{equation}\label{eq:deltaAction1_step2}
    \begin{aligned}
        - \delta I^{(1)}/N = \int_{\Omega} \Bigg\{ &
        \sum_x
        g_x^T(\Omega)
        \begin{pmatrix}
            0 & \frac{1}{2} \left(2 \zeta - J + i (-1)^x \Omega \right) \\
            \frac{1}{2} \left(2 \zeta - J - i (-1)^x \Omega \right) & 0 \\
        \end{pmatrix}
        g_x(-\Omega)
         \\&+ \sum_k \frac{-J}{2} g_k^T(\Omega) 
        \begin{pmatrix}
            0 & \cos{k} \\
            \cos{k} & 0 \\
        \end{pmatrix}
        g_k(-\Omega)
        \Bigg\}.
        \end{aligned}
    \end{equation}
    To deal with site dependent phase $(-1)^x$, we can define new enlarged basis $\mathfrak{g}_k^{ud} = \left( \delta G^{ud}_{k}, \delta G^{ud}_{k+\pi} \right)$, and  $\mathfrak{g}_k^{du} = \left( \delta G^{du}_{k}, \delta G^{du}_{k+\pi} \right)^T$. Focusing on $k\ll 1$, the action become
    \begin{equation}\label{eq:deltaAction1_step3}
        - \delta I^{(1)}/N =  \int_{\Omega, k\ll 1}
        \mathfrak{g}^{ud}_{k}(\Omega)
        \left(
        \begin{array}{cc}
            2 \zeta -J \cos (k)-J & -{i\Omega}   \\
             -{i \Omega}  & 2 \zeta +J \cos (k)-J 
        \end{array}
        \right)
        \mathfrak{g}^{du}_{-k}(-\Omega).
    \end{equation}  
    An expansion of \eqref{eq:deltaAction1_step3} leads to the result cited in the main text. We also compute the smaller eigenvalue of the matrix as
    \begin{equation}\label{eq:action_eigen}
        \begin{split}
            &\frac{1}{2} \left(4 \zeta -\sqrt{2} \sqrt{J^2 \cos (2 k)+J^2 - 2 \Omega ^2}-2 J\right) =\frac{\Omega^2}{2J} + \frac{J k^2}{2} + (2\zeta - 2J) .
        \end{split}
    \end{equation} 

    After integrating out $\mathfrak{g}_k^{du}$, the resulting $\delta$ function leads to two coupled constraint equations
    \begin{equation}
        \left(
        \begin{array}{cc}
            -2 \zeta +2J-\frac{Jk^2}{2} & {i\Omega}   \\
             {i \Omega}  & -2 \zeta +\frac{Jk^2}{2} 
        \end{array}
        \right)
        \begin{pmatrix}
            \delta G^{ud}_k \\
      \delta G^{ud}_{k+\pi} \\
        \end{pmatrix} = 0,
    \end{equation}
    which can be simplified as
    \begin{equation}
        \begin{split}
            &\left(-2 \zeta +2J-\frac{Jk^2}{2} + \frac{\Omega^2}{-2\zeta + Jk^2/2} \right) \delta G^{ud}_{k} = 0  \\
            \Rightarrow & - \left(2 \zeta - 2J + \frac{1}{2\zeta} \Omega^2 + \frac{Jk^2}{2} + \mathcal{O}(\Omega^2 k^2) \right) \delta G^{ud}_{k} = 0  \\
        \end{split}
    \end{equation}
    It leads to the differential equation in the main text. 
    \begin{equation}
     \left(-\frac{1}{2\zeta}\partial_t^2+2\zeta-2J\right) \delta G^{ud}_x(t,t)=0.
    \end{equation}  
    According to the translation invariant solution, the equal time Green's function in this case is
    \begin{equation}
        \begin{split}
            G_x(0^+) &= \left(
\begin{array}{cc}
 -\frac{1}{2} \left((-1)^x-1\right)  & 0 \\
 0 & -\frac{1}{2} \left((-1)^x+1\right)  \\
\end{array}
\right) \\
        G_x(0^-) &= \left(
\begin{array}{cc}
 -\frac{1}{2} \left((-1)^x+1\right)  & 0 \\
 0 & -\frac{1}{2} \left((-1)^x-1\right)  \\
\end{array}
\right) \\
        \end{split}
    \end{equation}
    Several observations help us to determine the detailed form of $\delta G^{ud}_k$. First, $\delta G^{ud}(t,t)$ is bounded by $G^{ud}(0^{\pm})=0$ when the fermion operator is away from the boundary, i.g. $t \ll T$. Therefore, $|\delta G^{ud}_x(t,t)| \sim e^{2t\sqrt{\zeta(\zeta-J)}}$ corresponds to the correct decaying direction.
    
    Second, in the main text, we have noticed the boundary condition $G_x^{du}(T,T)_\phi=G_x^{uu}(T,T)_\phi$ for $x\in \bar{A}$ and $G_x^{du}(T,T)_\phi=e^{i\phi}G_x^{uu}(T,T)_\phi$ for $x\in A$. Without loss of generality, we consider a special case where $x$ is at the boundary of $A$ , namely $x\in A,\text{ and } x+1 \in \bar{A}$. We can have $ |\delta G^{ud}_x(T+0^{-},T)| \sim |G^{uu}_x(0^{-})| = \frac{1}{2}\left| (-1)^x + 1\right|$ and $ |\delta G^{du}_{x+1}(T,T+0^{-})| \sim |G^{uu}_{x+1}(0^{+})| = \frac{1}{2}\left| (-1)^{x+1} - 1\right|$. If $x$ is even site, then we find both $\delta G^{ud}$ and $\delta G^{du}$ are $\mathcal{O}(1)$ value. Then we can write $|\delta G^{ud}_x(t,t)| \sim e^{-2(t-T)\sqrt{\zeta(\zeta-J)}}$ and $|\delta G^{ud}_x(t,t)| \sim e^{-2(t-T)\sqrt{\zeta(\zeta-J)}}$.

    Thirdly, we consider the phase difference between $\delta G^{ud}_A(t,t)$ and $\delta G^{du}_{\bar{A}}(t,t)$. As our previous discussion, the differences are a minus sign (referring to Eq.~\eqref{eq:saddlesolution}) and a $e^{i\phi}$ phase (referring to Eq.~(12), (13) in the main text). 
    \begin{equation}
        \begin{split}
            {J}N~\text{Re}\int dt~ G_{A}^{ud}(t,t)G_{\bar{A}}^{du}(t,t) 
    \sim {J}N~\text{Re} \int dt~ e^{-4(t-T)\sqrt{\zeta(\zeta-J)}} \left( -e^{i\phi} \right) 
      \sim -\frac{J}{\sqrt{\zeta(\zeta-J)}}N\cos\phi. \\
        \end{split}
    \end{equation}
    
    \subsection{$\zeta < J$}
    
    \paragraph{$\Tr \log term$}
    
    Taking saddle point Eq.~\eqref{eq:saddlesolution}, ~\eqref{eq:saddlesolutionzetasmall} and expanding the $\Tr \log$ term leads to 
    \begin{equation}\label{eq:M1_2_action}
        \begin{split}
            -\delta I_1^{(2)}/N =& - \frac12 \int_{\omega, \Omega} \Tr\left[ G_s(\omega + \Omega) \delta \Sigma(\Omega) G_s(\omega) \delta \Sigma(-\Omega) \right] \\
            =& \sum_x \int_{\Omega} \tilde{\sigma}^{T}_x(\Omega) \mathcal{M}_1^{(2)} \tilde{\sigma}_x(-\Omega),
        \end{split}
    \end{equation}
    where $\tilde{\sigma}_x = \left(\delta \Sigma^{uu}_x, \delta \Sigma^{dd}_x, \delta \Sigma^{ud}_x, \delta \Sigma^{du}_x \right)^T$. Since the off-diagonal coupling $G^{ud}_s, G^{du}_s$ is nonzero, the diagonal and off-diagonal parts of the self-energy are coupled. Therefore we consider all fluctuation components in the $\tilde{\sigma}_x$. The kernel reads
    \begin{equation}\label{eq:M1_2}
    \begin{aligned}
        &\mathcal{M}_1^{(2)}\sqrt{\zeta ^2+4 \mathcal{S}^2} \left(\zeta ^2+4 \mathcal{S}^2+\Omega ^2\right) =  \\&\left(
        \begin{array}{cccc}
            \mathcal{S}^2 & \mathcal{S}^2 & -\frac{1}{2} \mathcal{S} z \left(\zeta  (-1)^x+i \Omega \right) & \frac{\mathcal{S} \left(\zeta  (-1)^x-i \Omega \right)}{2 z} \\
            \mathcal{S}^2 & \mathcal{S}^2 & -\frac{1}{2} \mathcal{S} z \left(\zeta  (-1)^x+i \Omega \right) & \frac{\mathcal{S} \left(\zeta  (-1)^x-i \Omega \right)}{2 z} \\
            -\frac{1}{2} \mathcal{S} z \left(\zeta  (-1)^x-i \Omega \right) & -\frac{1}{2} \mathcal{S} z \left(\zeta  (-1)^x-i \Omega \right) & -\mathcal{S}^2 z^2 & \frac{1}{2} \left(-2 \mathcal{S}^2-\zeta  \left(\zeta -i (-1)^x \Omega \right)\right) \\
            \frac{\mathcal{S} \left(\zeta  (-1)^x+i \Omega \right)}{2 z} & \frac{\mathcal{S} \left(\zeta  (-1)^x+i \Omega \right)}{2 z} & \frac{1}{2} \left(-2 \mathcal{S}^2-\zeta  \left(\zeta +i (-1)^x \Omega \right)\right) & -\frac{\mathcal{S}^2}{z^2} \\
        \end{array}
        \right)
        \end{aligned}
    \end{equation}

    \paragraph{$\Sigma G$ term}
    For the $\Sigma G$ term, we also need to consider all components of fluctuated Green's function. Similarly, we defined $\tilde{g}_x = \left(\delta G^{uu}_x, \delta G^{dd}_x, \delta G^{ud}_x, \delta G^{du}_x \right)^T$. The linear coupling term reads that
    \begin{equation}\label{eq:SigmaG_2}
        \begin{split}
            &\int dtdt'~    \Sigma_x^{ac} (G_x^{ca})^T  \\
            =& -I_2^{(2)}/N +   \sum_x \frac{1}{2} \int_{\Omega} \left[ \tilde{g}_x^T(\Omega)  \begin{pmatrix}
            1 & 0 & 0 & 0 \\
            0 & 1 & 0 & 0 \\
            0 & 0 & 0 & 1 \\
            0 & 0 & 1 & 0 \\
            \end{pmatrix} \tilde{\sigma}_x(-\Omega) + \tilde{\sigma}^T_x(\Omega)
            \begin{pmatrix}
            1 & 0 & 0 & 0 \\
            0 & 1 & 0 & 0 \\
            0 & 0 & 0 & 1 \\
            0 & 0 & 1 & 0 \\
            \end{pmatrix} \tilde{g}_x(-\Omega) \right] \\
            \equiv&-I_2^{(2)}/N - \delta I_2^{(2)}/N.
        \end{split}
    \end{equation}
    
  \paragraph{$V$ term} In constrast to the $\zeta > J$ case, the second order off-diagonal term contributed by $V$ term is non-zero. Together with the diagonal term, we calculate the second-order fluctuation action 
    \begin{equation}\label{eq:V_action2}
        \begin{split}
            &\int \diff t \diff t' \sum_x
            f^a f^c \hat{I}   \frac{V}{4} (G_x^{ac})^2 \left((-G_x^{ca})^2\right)^T \\
            =& -I_3^{(2)}/N +  \frac{1}{4} \Big(-6 \delta G^{uu}_x(-\Omega) \delta G^{uu}_x(\Omega)
             G^{uu}_{s,x}(t=0)^2
             -6 \delta G^{dd}_x(-\Omega) \delta G^{dd}_x(\Omega) G^{dd}_{s,x}(t=0)^2  \\
             & + 4 G^{ud}_{s,x}(t=0) G^{du}_{s,x}(t=0) (\delta G^{ud}_x(-\Omega) \delta G^{du}_x(\Omega)+\delta G^{du}_x(-\Omega) \delta G^{ud}_x(\Omega)) \\
             &+2 \delta G^{ud}_x(-\Omega) \delta G^{ud}_x(\Omega) G^{du}_{s,x}(t=0)^2+2 \delta G^{du}_x(-\Omega) \delta G^{du}_x(\Omega) G^{ud}_{s,x}(t=0)^2\Big)  \\
             =& -I_3^{(2)}/N +  \sum_x \int_{\Omega} \tilde{g}^{T}_x(\Omega) \mathcal{M}_3^{(2)} \tilde{g}_x(-\Omega) \\
             \equiv& -I_3^{(2)}/N - \delta I_3^{(2)}/N, \\
        \end{split}
    \end{equation}
    where the kernel reads that
    \begin{equation}\label{eq:M3_2}
        \mathcal{M}_3^{(2)} = \frac{1}{\zeta ^2+4 \mathcal{S}^2} \left(
        \begin{array}{cccc}
            \frac{-3 \zeta  (-1)^x z^2 \sqrt{\zeta ^2+4 \mathcal{S}^2}-6 \mathcal{S}^2 z^2-3 \zeta ^2 z^2}{4 z^2} & 0 & 0 & 0 \\
            0 & \frac{3 \zeta  (-1)^x z^2 \sqrt{\zeta ^2+4 \mathcal{S}^2}-6 \mathcal{S}^2 z^2-3 \zeta ^2 z^2}{4 z^2} & 0 & 0 \\
            0 & 0 & \frac{\mathcal{S}^2 z^2}{2} & -\mathcal{S}^2 \\
            0 & 0 & -\mathcal{S}^2 & \frac{\mathcal{S}^2}{2 z^2} \\
        \end{array}
        \right).
    \end{equation}

 \paragraph{$J$ term}
\begin{equation}\label{eq:J_action2}
    \begin{split}
        &\int \diff t \diff t' \sum_x
        \frac{J}{4} \left[G^{ac}_{x+1} (-G^{ca}_{x})^T + G^{ac}_{x-1} (-G^{ca}_{x})^T\right] \\
        =& - I_4^{(1)}/N + \int_{k,\Omega}
        \frac{J}{2} \tilde{g}_k^T(\Omega) 
        \begin{pmatrix}
            \cos{k} & 0 &   0 & 0 \\
            0 & \cos{k} &   0 & 0 \\
            0 & 0 & 0 & -\cos{k} \\
            0 & 0 & -\cos{k} & 0 \\
        \end{pmatrix}
        \tilde{g}_{-k}(-\Omega)\\
        \equiv& - I_4^{(1)}/N - \delta I_4^{(1)}/N, \\
    \end{split}
\end{equation}

    \paragraph{Effective Action} To obtain the effective action, we need to integrate out self-energy $\tilde{\sigma}$. However, we find there are two zero eigenvalues in the kernel of fluctuated self-energy \eqref{eq:M1_2}. Since self-energy $\delta \Sigma$ and Green's function $\delta G$ are linearly coupled, these two zero mode lead to two constraints of $\delta G$. After applying these constraints, we can safely integrate out self-energy in the reduced subspace. Finally, we expect to obtain effective action with $4 - 2 = 2$ fluctuation fields in the $x$ space.
    
    We perform the calculations in detail following the arguments above. First, the four eigenvalues of $\mathcal{M}_1^{(2)}$ is 
    \begin{equation}\label{eq:M1_2_eigenvals}
        \begin{split}
            \lambda_{1,2} &= 0 \\
            \lambda_{3} & = \frac{\sqrt{\Omega ^2 z^2 \left(2 \mathcal{S}^2 \left(z^4+1\right)+\zeta ^2 z^2\right)+\left(\mathcal{S}^2 \left(z^2+1\right)^2+\zeta ^2 z^2\right)^2}-\mathcal{S}^2 \left(z^2-1\right)^2}{2 z^2 \sqrt{\zeta ^2+4 \mathcal{S}^2} \left(\zeta ^2+4 \mathcal{S}^2+\Omega ^2\right)} \\
            \lambda_{4}&=\frac{-\sqrt{\Omega ^2 z^2 \left(2 \mathcal{S}^2 \left(z^4+1\right)+\zeta ^2 z^2\right)+\left(\mathcal{S}^2 \left(z^2+1\right)^2+\zeta ^2 z^2\right)^2}-\mathcal{S}^2 \left(z^2-1\right)^2}{2 z^2 \sqrt{\zeta ^2+4 \mathcal{S}^2} \left(\zeta ^2+4 \mathcal{S}^2+\Omega ^2\right)}. \\
        \end{split}
    \end{equation}  
    The corresponding eigenvectors without normalization are
    \begin{equation}\label{eq:M1_2_eigenvecs}
        \begin{split}
            u_1 &= (1,-1,0,0)^T \\
            u_2 &= \left(0,\frac{\zeta  (-1)^{x+1}}{S z},-\frac{1}{z^2},1\right)^T \\
            u_3 &= \left(
            \begin{array}{c}
                \frac{S \left(\sqrt{\Omega ^2 z^2 \left(2 S^2 \left(z^4+1\right)+\zeta ^2 z^2\right)+\left(S^2 \left(z^2+1\right)^2+\zeta ^2 z^2\right)^2}+S^2 \left(z^2+1\right)^2+\zeta  z^4 \left(\zeta +i (-1)^x \Omega \right)\right)}{2 S^2 \left(\zeta  (-1)^x \left(z^3+z\right)+i \Omega  z\right)+\zeta ^2 z^3 \left(\zeta  (-1)^x+i \Omega \right)} \\
                \frac{S \left(\sqrt{\Omega ^2 z^2 \left(2 S^2 \left(z^4+1\right)+\zeta ^2 z^2\right)+\left(S^2 \left(z^2+1\right)^2+\zeta ^2 z^2\right)^2}+S^2 \left(z^2+1\right)^2+\zeta  z^4 \left(\zeta +i (-1)^x \Omega \right)\right)}{2 S^2 \left(\zeta  (-1)^x \left(z^3+z\right)+i \Omega  z\right)+\zeta ^2 z^3 \left(\zeta  (-1)^x+i \Omega \right)} \\
                \frac{\zeta  (-1)^{x+1} \sqrt{\Omega ^2 z^2 \left(2 S^2 \left(z^4+1\right)+\zeta ^2 z^2\right)+\left(S^2 \left(z^2+1\right)^2+\zeta ^2 z^2\right)^2}+S^2 \left(\zeta  (-1)^x \left(z^4-1\right)+2 i \Omega  z^2\right)}{2 S^2 \left(\zeta  (-1)^x \left(z^2+1\right)+i \Omega \right)+\zeta ^2 z^2 \left(\zeta  (-1)^x+i \Omega \right)} \\
                1 \\
            \end{array}
            \right) \\
            u_4 &= \left(
            \begin{array}{c}
                \frac{S \left(-\sqrt{\Omega ^2 z^2 \left(2 S^2 \left(z^4+1\right)+\zeta ^2 z^2\right)+\left(S^2 \left(z^2+1\right)^2+\zeta ^2 z^2\right)^2}+S^2 \left(z^2+1\right)^2+\zeta  z^4 \left(\zeta +i (-1)^x \Omega \right)\right)}{2 S^2 \left(\zeta  (-1)^x \left(z^3+z\right)+i \Omega  z\right)+\zeta ^2 z^3 \left(\zeta  (-1)^x+i \Omega \right)} \\
                \frac{S \left(-\sqrt{\Omega ^2 z^2 \left(2 S^2 \left(z^4+1\right)+\zeta ^2 z^2\right)+\left(S^2 \left(z^2+1\right)^2+\zeta ^2 z^2\right)^2}+S^2 \left(z^2+1\right)^2+\zeta  z^4 \left(\zeta +i (-1)^x \Omega \right)\right)}{2 S^2 \left(\zeta  (-1)^x \left(z^3+z\right)+i \Omega  z\right)+\zeta ^2 z^3 \left(\zeta  (-1)^x+i \Omega \right)} \\
                \frac{\zeta  (-1)^x \sqrt{\Omega ^2 z^2 \left(2 S^2 \left(z^4+1\right)+\zeta ^2 z^2\right)+\left(S^2 \left(z^2+1\right)^2+\zeta ^2 z^2\right)^2}+S^2 \left(\zeta  (-1)^x \left(z^4-1\right)+2 i \Omega  z^2\right)}{2 S^2 \left(\zeta  (-1)^x \left(z^2+1\right)+i \Omega \right)+\zeta ^2 z^2 \left(\zeta  (-1)^x+i \Omega \right)} \\
                1 \\
            \end{array}
            \right). \\
        \end{split}
    \end{equation}
    The eigenvalues and eigenvectors satisfy $\mathcal{M}_1 U = U \textbf{Diag}\{\lambda_1,\lambda_2,\lambda_3,\lambda_4\}$, where $U=\left(\frac{u_1}{\norm{u_1}},\frac{u_2}{\norm{u_2}},\frac{u_3}{\norm{u_3}},\frac{u_4}{\norm{u_4}} \right)$ and $\textbf{Diag}$ means to construct diagonal matrix using the elements. The transformation matrix $U$ is normalized, and therefore satisfies the unitary condition $U U^{\dagger} = \hat{1}$. We consider the $\Sigma G$ and $\Tr \log$ term
    \begin{equation}\label{eq:integrate_2}
    \begin{aligned}
            &\sum_x \int_{\Omega} \tilde{\sigma}^{T}_x(\Omega) U \textbf{Diag}\{\lambda_1,\lambda_2,\lambda_3,\lambda_4\} U^{\dagger} \tilde{\sigma}_x(-\Omega) \\&+ 
            \sum_x \frac{1}{2} \int_{\Omega} \left[ \tilde{g}_x^T(\Omega)  \begin{pmatrix}
                1 & 0 & 0 & 0 \\
                0 & 1 & 0 & 0 \\
                0 & 0 & 0 & 1 \\
                0 & 0 & 1 & 0 \\
            \end{pmatrix} U U^{\dagger} \tilde{\sigma}_x(-\Omega) + \tilde{\sigma}^T_x(\Omega) U U^{\dagger}
            \begin{pmatrix}
                1 & 0 & 0 & 0 \\
                0 & 1 & 0 & 0 \\
                0 & 0 & 0 & 1 \\
                0 & 0 & 1 & 0 \\
            \end{pmatrix} \tilde{g}_x(-\Omega) \right] \\
        & = \sum_x \int_{\Omega} \tilde{\sigma}^{\dagger}_{\text{eig},x}(\Omega) \textbf{Diag}\{\lambda_1,\lambda_2,\lambda_3,\lambda_4\}  \tilde{\sigma}_{\text{eig},x}(-\Omega) + 
        \sum_x \frac{1}{2} \int_{\Omega} \left[ \tilde{g}_{\text{eig},x}^{\dagger}(\Omega)   \tilde{\sigma}_{\text{eig},x}(-\Omega) + \tilde{\sigma}^{\dagger}_{\text{eig},x}(\Omega)  \tilde{g}_{\text{eig},x}(-\Omega) \right] \\
        \end{aligned}
    \end{equation}
    Here we define the field in the eigenbasis 
    \begin{equation}\label{eq:define_eig}
        \begin{split}
            & \tilde{\sigma}_{\text{eig},x} = U^{\dagger} \tilde{\sigma}_{x}, \tilde{\sigma}_{\text{eig},x}^{\dagger} = \tilde{\sigma}_{x}^T U\\
            & \tilde{g}_{\text{eig},x}^{\dagger}(\Omega)  = \tilde{g}_x^T(\Omega)  \begin{pmatrix}
                1 & 0 & 0 & 0 \\
                0 & 1 & 0 & 0 \\
                0 & 0 & 0 & 1 \\
                0 & 0 & 1 & 0 \\
            \end{pmatrix} U  \\
        & \tilde{g}_{\text{eig},x}(-\Omega)  = U^{\dagger}
        \begin{pmatrix}
            1 & 0 & 0 & 0 \\
            0 & 1 & 0 & 0 \\
            0 & 0 & 0 & 1 \\
            0 & 0 & 1 & 0 \\
        \end{pmatrix} \tilde{g}_x(-\Omega) \\
        \end{split}
    \end{equation}
     Since $\lambda_{1,2}=0$, the absence of quadratic $(\tilde{\sigma}_{\text{eig},x})_1, (\tilde{\sigma}_{\text{eig},x})_2$ leads to two delta function $\delta((\tilde{g}_{\text{eig},x})_1), \delta((\tilde{g}_{\text{eig},x})_2) $ when performing the integration on self-energy, which leads to two constraints in the Green's function. In detail,
     \begin{equation}\label{eq:constraint}
        \begin{split}
            (\tilde{g}_{\text{eig},x}^{\dagger})_1 &= \left(\delta G^{uu}_x, \delta G^{dd}_x, \delta G^{du}_x, \delta G^{ud}_x \right) u_1 \propto -\delta G^{uu}_x + \delta G^{dd}_x = 0\\
            (\tilde{g}_{\text{eig},x}^{\dagger})_2 &= \left(\delta G^{uu}_x, \delta G^{dd}_x, \delta G^{du}_x, \delta G^{ud}_x \right) u_2 \propto \frac{\zeta  (-1)^{x+1}}{\mathcal{S} z}\delta G^{dd}_x  -\frac{1}{z^2}\delta G^{du}_x + \delta G^{ud}_x = 0.\\
        \end{split}
     \end{equation}
    These two constraints reduce to $\delta G^{uu}_x = \delta G^{dd}_x = 0$ when we set $\mathcal{S}=0$. It is consistent with the $\zeta > J$ case where the diagonal terms are decoupled from the off-diagonal terms and therefore to be ignored. Returning to the Eq.~\eqref{eq:integrate_2}, we can integrate out the residual self-energy
    \begin{equation}\label{eq:integrate_2_step2}
        \begin{aligned}
            &  \sum_x \int_{\Omega} \tilde{\sigma}^{\dagger}_{\text{eig},x}(\Omega) \textbf{Diag}\{\lambda_1,\lambda_2,\lambda_3,\lambda_4\}  \tilde{\sigma}_{\text{eig},x}(-\Omega) + 
            \sum_x \frac{1}{2} \int_{\Omega} \left[ \tilde{g}_{\text{eig},x}^{\dagger}(\Omega)   \tilde{\sigma}_{\text{eig},x}(-\Omega) + \tilde{\sigma}^{\dagger}_{\text{eig},x}(\Omega)  \tilde{g}_{\text{eig},x}(-\Omega) \right] \\
            \Rightarrow&  \sum_x \int_{\Omega} \tilde{\sigma}^{\dagger}_{\text{eig,R},x}(\Omega) \textbf{Diag}\{\lambda_3,\lambda_4\}  \tilde{\sigma}_{\text{eig,R},x}(-\Omega) + 
            \sum_x \frac{1}{2} \int_{\Omega} \left[ \tilde{g}_{\text{eig,R},x}^T(\Omega)   \tilde{\sigma}_{\text{eig,R},x}(-\Omega) + \tilde{\sigma}^{\dagger}_{\text{eig,R},x}(\Omega)  \tilde{g}_{\text{eig,R},x}(-\Omega) \right] \\
            \Rightarrow&  -\frac{1}{4} \sum_x \int_{\Omega} \tilde{g}^{\dagger}_{\text{eig,R},x}(\Omega) \textbf{Diag}\{\lambda_3^{-1},\lambda_4^{-1}\}  \tilde{g}_{\text{eig,R},x}(-\Omega).  \\
        \end{aligned}
    \end{equation}
    Here the $\text{R}$ subscript means to take 3, 4 components in the eigenbasis, according to the definition in Eq.~\eqref{eq:define_eig}. Except $\Sigma G$ and $\Tr \log$ term, we consider $V$ term and $J$ term. We apply constraint Eq.~\eqref{eq:constraint} in Eq.~\eqref{eq:V_action2}, \eqref{eq:J_action2} correspondingly. Then we sum all contributions together and simplify the result, which leads to
    \begin{equation}\label{eq:action2_simplify}
        \begin{split}
            -\delta I^{(2)} /N &= \sum_x  \int_{\Omega} g_x^T(\Omega)
            \left(
            \begin{array}{cc}
                -\frac{S^2 z^2 \sqrt{\zeta ^2+4 S^2}}{\zeta ^2} & \frac{\sqrt{\zeta ^2+4 S^2} \left(2 S^2+\zeta  (\zeta +(-1)^x i \Omega )\right)}{2 \zeta ^2} \\
                \frac{\sqrt{\zeta ^2+4 S^2} \left(2 S^2+\zeta  (\zeta - (-1)^x i \Omega )\right)}{2 \zeta ^2} & -\frac{S^2 \sqrt{\zeta ^2+4 S^2}}{\zeta ^2 z^2} \\
            \end{array}
            \right) g_x(-\Omega) \\
            &+ V \sum_x  \int_{\Omega} g_x^T(\Omega)
            \left(
            \begin{array}{cc}
                -\frac{S^2 z^2 \left(\zeta ^2+3 S^2\right)}{\zeta ^4+4 \zeta ^2 S^2} & \frac{\zeta ^2 S^2+6 S^4}{2 \zeta ^4+8 \zeta ^2 S^2} \\
                \frac{\zeta ^2 S^2+6 S^4}{2 \zeta ^4+8 \zeta ^2 S^2} & -\frac{S^2 \left(\zeta ^2+3 S^2\right)}{\zeta ^2 z^2 \left(\zeta ^2+4 S^2\right)} \\
            \end{array}
            \right)
             g_x(-\Omega) \\
             &+ \frac{J}{2} \sum_k  \int_{\Omega} g_{k}^T(\Omega)
             \left(
             \begin{array}{cc}
                \frac{2 S^2 z^2 \cos (k)}{\zeta ^2} & \cos (k) \left(-\frac{2 S^2}{\zeta ^2}-1\right) \\
                \cos (k) \left(-\frac{2 S^2}{\zeta ^2}-1\right) & \frac{2 S^2 \cos (k)}{\zeta ^2 z^2} \\
             \end{array}
             \right)
             g_{-k}(-\Omega). \\
        \end{split}
    \end{equation}
    Here the effective action is still in the bilinear form of vector $g_x = \left( \delta G^{ud}_x, \delta G^{du}_x \right)^T$, since we have applied two constraints in the derivation. As we point in the main text, we need two identifications
    \begin{equation}\label{eq:action2_identification}
        \begin{split}
            \left(\delta G^{ud}_x(t,t), \delta G^{du}_x(t,t) \right)&= \varphi_1(x,t) \left(-iG^{ud}(0),  iG^{du}(0)\right) \\
            \left(\delta G^{ud}_x(t,t), \delta G^{du}_x(t,t) \right)&= \varphi_2(x,t) \left(-iG^{du}(0),  -iG^{ud}(0)\right). \\
        \end{split}
    \end{equation}
    With these two identifications, the action is projected on two new subspaces, where
    \begin{equation}\label{eq:action2_phi}
        \begin{aligned}
                -\frac{\delta I^{(2)} }{N} &= \int_{\Omega, k}  (\varphi_1,\varphi_2)
            \left(
            \begin{array}{cc}
                \frac{J \mathcal{S}^2 (\cos(k) -1)}{\zeta ^2+4 \mathcal{S}^2} & -\frac{J \mathcal{S}^2 \left(z^4-1\right) (\cos(k) -1)}{2 z^2 \left(\zeta ^2+4 \mathcal{S}^2\right)} \\
                -\frac{J \mathcal{S}^2 \left(z^4-1\right) (\cos(k) -1)}{2 z^2 \left(\zeta ^2+4 \mathcal{S}^2\right)} & K_{2,2}^\varphi \\
            \end{array}
            \right)
            \begin{pmatrix}
                \varphi_1 \\
                \varphi_2 \\
            \end{pmatrix} \\
        +& \int_{\Omega} \sum_x  (\varphi_1,\varphi_2)(x,\Omega)
        \left(
        \begin{array}{cc}
            0 & -\frac{i \mathcal{S}^2 (-1)^x \Omega  \left(z^4+1\right)}{2 \zeta  z^2 \sqrt{\zeta ^2+4 \mathcal{S}^2}} \\
            \frac{i \mathcal{S}^2 (-1)^x \Omega  \left(z^4+1\right)}{2 \zeta  z^2 \sqrt{\zeta ^2+4 \mathcal{S}^2}} & 0 \\
        \end{array}
        \right)
        \begin{pmatrix}
            \varphi_1 \\
            \varphi_2 \\
        \end{pmatrix}(x,-\Omega), \\
        \end{aligned}   
    \end{equation}
    with 
    \begin{equation}
    K_{2,2}^\varphi=\frac{\mathcal{S}^2 \left(-J \mathcal{S}^2 \left(z^4+1\right)^2 (\cos(k) +3)-\zeta ^2 J \left(z^4 (\cos(k) +1)+z^8+1\right)+\left(z^4+1\right)^2 \left(\zeta ^2+4 \mathcal{S}^2\right)^{3/2}\right)}{\zeta ^2 z^4 \left(\zeta ^2+4 \mathcal{S}^2\right)}.
    \end{equation}
    Here we have used the identity of parameters Eq.~\eqref{eq:saddlesolutionzetasmall}. Finally, similar to the $\zeta > J$ subsection, we rewrite the action in the enlarged basis $\bm{\varphi}(k,\Omega) =\left( \varphi_1(k,\Omega), \varphi_1(k+\pi,\Omega), \varphi_2(k,\Omega), \varphi_2(k+\pi,\Omega) \right)^T$, which reads that
    \begin{equation}\label{eq:action2_phi_enlarge}
        -\frac{\delta I^{(2)} }{N}= \int_{\Omega, k}  \bm{\varphi}(k,\Omega)
         \mathcal{M}^{(2)}
        \bm{\varphi}(-k,-\Omega),
    \end{equation}
    where the kernel is 
\begin{equation}\label{eq:M2}
    \mathcal{M}^{(2)}=
    \begin{pmatrix}
        \frac{J \mathcal{S}^2 ( \cos(k) -1)}{2 \left(\zeta ^2+4 \mathcal{S}^2\right)} & \bm{v} \\
        \bm{v}^{\dagger} & \mathcal{M}^{(2)}_{\text{gap}}
    \end{pmatrix},
\end{equation}
where 
\begin{equation}\label{eq:v}
    \begin{split}
        \bm{v} &= \begin{pmatrix}
            0 & -\frac{J \mathcal{S}^2 \left(z^4-1\right) (\cos (k)-1)}{4 z^2 \left(\zeta ^2+4 \mathcal{S}^2\right)} & -\frac{i \mathcal{S}^2 \Omega  \left(z^4+1\right)}{4 \zeta  z^2 \sqrt{\zeta ^2+4 \mathcal{S}^2}} \\
        \end{pmatrix}, \\
    \end{split}
\end{equation}
    \begin{equation}\label{eq:Mgap}
        \mathcal{M}^{(2)}_{\text{gap}} = 
        \left(
        \begin{array}{ccc}
            -\frac{J \mathcal{S}^2 (\cos (k)+1)}{2 \left(\zeta ^2+4 \mathcal{S}^2\right)} & -\frac{i \mathcal{S}^2 \Omega  \left(z^4+1\right)}{4 \zeta  z^2 \sqrt{\zeta ^2+4 \mathcal{S}^2}} & \frac{J \mathcal{S}^2 \left(z^4-1\right) (\cos (k)+1)}{4 z^2 \left(\zeta ^2+4 \mathcal{S}^2\right)} \\
            \frac{i \mathcal{S}^2 \Omega  \left(z^4+1\right)}{4 \zeta  z^2 \sqrt{\zeta ^2+4 \mathcal{S}^2}} & (\mathcal{M}^{(2)}_{\text{gap}})_{2,2} & 0 \\
            \frac{J \mathcal{S}^2 \left(z^4-1\right) (\cos (k)+1)}{4 z^2 \left(\zeta ^2+4 \mathcal{S}^2\right)} & 0 & (\mathcal{M}^{(2)}_{\text{gap}})_{3,3} \\
        \end{array}
        \right) ,
    \end{equation}
    with 
\begin{equation}
\begin{aligned}
(\mathcal{M}^{(2)}_{\text{gap}})_{2,2}&=\frac{\mathcal{S}^2 \left(-J \cos (k) \left(\mathcal{S}^2 \left(z^4+1\right)^2+\zeta ^2 z^4\right)-3 J \mathcal{S}^2 \left(z^4+1\right)^2-\zeta ^2 J \left(z^8+z^4+1\right)+\left(z^4+1\right)^2 \left(\zeta ^2+4 \mathcal{S}^2\right)^{3/2}\right)}{2 \zeta ^2 z^4 \left(\zeta ^2+4 \mathcal{S}^2\right)},\\
(\mathcal{M}^{(2)}_{\text{gap}})_{3,3}&=\frac{\mathcal{S}^2 \left(J \cos (k) \left(\mathcal{S}^2 \left(z^4+1\right)^2+\zeta ^2 z^4\right)-3 J \mathcal{S}^2 \left(z^4+1\right)^2-\zeta ^2 J \left(z^8+z^4+1\right)+\left(z^4+1\right)^2 \left(\zeta ^2+4 \mathcal{S}^2\right)^{3/2}\right)}{2 \zeta ^2 z^4 \left(\zeta ^2+4 \mathcal{S}^2\right)}.
\end{aligned}
\end{equation}
We derive the final action in $k\to0, \Omega\to0$ limit. We find $\varphi_1(k)$ corresponds to a gapless mode and the others field correspond to gapped modes. Therefore we can integrate out the gapped modes to obtain effective action. In detail, we keep the $\bm{v}$ to the leading order
\begin{equation}\label{eq:v_leadingorder}
    \tilde{\bm{v}} = \left(
    \begin{array}{ccc}
        0 & \frac{J k^2 \mathcal{S}^2 \left(z^4-1\right)}{8 z^2 \left(\zeta ^2+4 \mathcal{S}^2\right)} & -\frac{i \mathcal{S}^2 \Omega  \left(z^4+1\right)}{4 \zeta  z^2 \sqrt{\zeta ^2+4 \mathcal{S}^2}} \\
    \end{array}
    \right),
\end{equation}
and take $\lim_{k\to0, \Omega\to0} \mathcal{M}^{(2)}_{\text{gap}}$. The final gapless mode can be obtained
\begin{equation}\label{eq:action2_integrate_gapped_mode}
    \begin{split}
            -\delta I^{(2)} /N &= \int_{\Omega, k}  \varphi_1(k,\Omega)
        \left[ -\frac{J k^2 \mathcal{S}^2}{4 \left(\zeta ^2+4 \mathcal{S}^2\right)} - \tilde{\bm{v}} \left(\lim_{k\to0, \Omega\to0} \mathcal{M}^{(2)}_{\text{gap}}\right)^{-1} \tilde{\bm{v}}^{\dagger} \right] 
        \varphi_1(-k,-\Omega) \\
        &\approxeq \int_{\Omega, k}  \varphi_1(k,\Omega)
        \left[ -\frac{J k^2 \mathcal{S}^2}{4 \left(\zeta ^2+4 \mathcal{S}^2\right)} - \frac{\mathcal{S}^2 \Omega ^2}{4 \left(\zeta ^2+4 \mathcal{S}^2\right) \left(2 \sqrt{\zeta ^2+4 \mathcal{S}^2}-J\right)} \right] 
        \varphi_1(-k,-\Omega) \\
        &\approxeq \int_{\Omega, k}  \varphi_1(x,\Omega)
        \left[ -\frac{J k^2 \mathcal{S}^2}{4 \left(\zeta ^2+4 \mathcal{S}^2\right)} - \frac{\mathcal{S}^2 \Omega ^2}{4 \left(\zeta ^2+4 \mathcal{S}^2\right) \left(2 \sqrt{\zeta ^2+4 \mathcal{S}^2}-J\right)} \right] 
        \varphi_1(-k,-\Omega). \\
    \end{split}
\end{equation}
In the leading order, we drop the high order $k^4$ term. Finally, the effective action reads that
\begin{equation}\label{eq:action2_main}
        \begin{split}
                S_{\text{eff}}&=\frac{N\mathcal{S}^2}{4(\zeta ^2+4 \mathcal{S}^2)}\int_{x,t}~\Bigg(\frac{1}{2 \sqrt{\zeta ^2+4 \mathcal{S}^2}-J}(\partial_t\varphi)^2  +J(\partial_x\varphi)^2\Bigg). \\
        \end{split}
    \end{equation}
According to Eq.~\eqref{eq:saddlesolutionzetasmall}, we have $\sqrt{\zeta ^2+4 S^2}\geq J$, and therefore the coefficients of the gapless mode are always positive.

\twocolumngrid
\bibliography{main.bbl}

\begin{thebibliography}{65}%
\makeatletter
\providecommand \@ifxundefined [1]{%
 \@ifx{#1\undefined}
}%
\providecommand \@ifnum [1]{%
 \ifnum #1\expandafter \@firstoftwo
 \else \expandafter \@secondoftwo
 \fi
}%
\providecommand \@ifx [1]{%
 \ifx #1\expandafter \@firstoftwo
 \else \expandafter \@secondoftwo
 \fi
}%
\providecommand \natexlab [1]{#1}%
\providecommand \enquote  [1]{``#1''}%
\providecommand \bibnamefont  [1]{#1}%
\providecommand \bibfnamefont [1]{#1}%
\providecommand \citenamefont [1]{#1}%
\providecommand \href@noop [0]{\@secondoftwo}%
\providecommand \href [0]{\begingroup \@sanitize@url \@href}%
\providecommand \@href[1]{\@@startlink{#1}\@@href}%
\providecommand \@@href[1]{\endgroup#1\@@endlink}%
\providecommand \@sanitize@url [0]{\catcode `\\12\catcode `\$12\catcode
  `\&12\catcode `\#12\catcode `\^12\catcode `\_12\catcode `\%12\relax}%
\providecommand \@@startlink[1]{}%
\providecommand \@@endlink[0]{}%
\providecommand \url  [0]{\begingroup\@sanitize@url \@url }%
\providecommand \@url [1]{\endgroup\@href {#1}{\urlprefix }}%
\providecommand \urlprefix  [0]{URL }%
\providecommand \Eprint [0]{\href }%
\providecommand \doibase [0]{https://doi.org/}%
\providecommand \selectlanguage [0]{\@gobble}%
\providecommand \bibinfo  [0]{\@secondoftwo}%
\providecommand \bibfield  [0]{\@secondoftwo}%
\providecommand \translation [1]{[#1]}%
\providecommand \BibitemOpen [0]{}%
\providecommand \bibitemStop [0]{}%
\providecommand \bibitemNoStop [0]{.\EOS\space}%
\providecommand \EOS [0]{\spacefactor3000\relax}%
\providecommand \BibitemShut  [1]{\csname bibitem#1\endcsname}%
\let\auto@bib@innerbib\@empty
\bibitem [{\citenamefont {Mazzucchi}\ \emph {et~al.}(2016)\citenamefont
  {Mazzucchi}, \citenamefont {Kozlowski}, \citenamefont {{Caballero-Benitez}},
  \citenamefont {Elliott},\ and\ \citenamefont
  {Mekhov}}]{mazzucchiQuantumMeasurementinducedDynamics2016}%
  \BibitemOpen
  \bibfield  {author} {\bibinfo {author} {\bibfnamefont {G.}~\bibnamefont
  {Mazzucchi}}, \bibinfo {author} {\bibfnamefont {W.}~\bibnamefont
  {Kozlowski}}, \bibinfo {author} {\bibfnamefont {S.~F.}\ \bibnamefont
  {{Caballero-Benitez}}}, \bibinfo {author} {\bibfnamefont {T.~J.}\
  \bibnamefont {Elliott}},\ and\ \bibinfo {author} {\bibfnamefont {I.~B.}\
  \bibnamefont {Mekhov}},\ }\href {https://doi.org/10.1103/PhysRevA.93.023632}
  {\bibfield  {journal} {\bibinfo  {journal} {Phys. Rev. A}\ }\textbf {\bibinfo
  {volume} {93}},\ \bibinfo {pages} {023632} (\bibinfo {year}
  {2016})}\BibitemShut {NoStop}%
\bibitem [{\citenamefont {Li}\ \emph {et~al.}(2018)\citenamefont {Li},
  \citenamefont {Chen},\ and\ \citenamefont
  {Fisher}}]{liQuantumZenoEffect2018}%
  \BibitemOpen
  \bibfield  {author} {\bibinfo {author} {\bibfnamefont {Y.}~\bibnamefont
  {Li}}, \bibinfo {author} {\bibfnamefont {X.}~\bibnamefont {Chen}},\ and\
  \bibinfo {author} {\bibfnamefont {M.~P.~A.}\ \bibnamefont {Fisher}},\ }\href
  {https://doi.org/10.1103/PhysRevB.98.205136} {\bibfield  {journal} {\bibinfo
  {journal} {Phys. Rev. B}\ }\textbf {\bibinfo {volume} {98}},\ \bibinfo
  {pages} {205136} (\bibinfo {year} {2018})}\BibitemShut {NoStop}%
\bibitem [{\citenamefont {Skinner}\ \emph {et~al.}(2019)\citenamefont
  {Skinner}, \citenamefont {Ruhman},\ and\ \citenamefont
  {Nahum}}]{skinnerMeasurementInducedPhaseTransitions2019}%
  \BibitemOpen
  \bibfield  {author} {\bibinfo {author} {\bibfnamefont {B.}~\bibnamefont
  {Skinner}}, \bibinfo {author} {\bibfnamefont {J.}~\bibnamefont {Ruhman}},\
  and\ \bibinfo {author} {\bibfnamefont {A.}~\bibnamefont {Nahum}},\ }\href
  {https://doi.org/10.1103/PhysRevX.9.031009} {\bibfield  {journal} {\bibinfo
  {journal} {Phys. Rev. X}\ }\textbf {\bibinfo {volume} {9}},\ \bibinfo {pages}
  {031009} (\bibinfo {year} {2019})}\BibitemShut {NoStop}%
\bibitem [{\citenamefont {Li}\ \emph {et~al.}(2019)\citenamefont {Li},
  \citenamefont {Chen},\ and\ \citenamefont
  {Fisher}}]{liMeasurementdrivenEntanglementTransition2019}%
  \BibitemOpen
  \bibfield  {author} {\bibinfo {author} {\bibfnamefont {Y.}~\bibnamefont
  {Li}}, \bibinfo {author} {\bibfnamefont {X.}~\bibnamefont {Chen}},\ and\
  \bibinfo {author} {\bibfnamefont {M.~P.~A.}\ \bibnamefont {Fisher}},\ }\href
  {https://doi.org/10.1103/PhysRevB.100.134306} {\bibfield  {journal} {\bibinfo
   {journal} {Phys. Rev. B}\ }\textbf {\bibinfo {volume} {100}},\ \bibinfo
  {pages} {134306} (\bibinfo {year} {2019})}\BibitemShut {NoStop}%
\bibitem [{\citenamefont {Szyniszewski}\ \emph {et~al.}(2019)\citenamefont
  {Szyniszewski}, \citenamefont {Romito},\ and\ \citenamefont
  {Schomerus}}]{szyniszewskiEntanglementTransitionVariablestrength2019}%
  \BibitemOpen
  \bibfield  {author} {\bibinfo {author} {\bibfnamefont {M.}~\bibnamefont
  {Szyniszewski}}, \bibinfo {author} {\bibfnamefont {A.}~\bibnamefont
  {Romito}},\ and\ \bibinfo {author} {\bibfnamefont {H.}~\bibnamefont
  {Schomerus}},\ }\href {https://doi.org/10.1103/PhysRevB.100.064204}
  {\bibfield  {journal} {\bibinfo  {journal} {Phys. Rev. B}\ }\textbf {\bibinfo
  {volume} {100}},\ \bibinfo {pages} {064204} (\bibinfo {year}
  {2019})}\BibitemShut {NoStop}%
\bibitem [{\citenamefont {Chan}\ \emph {et~al.}(2019)\citenamefont {Chan},
  \citenamefont {Nandkishore}, \citenamefont {Pretko},\ and\ \citenamefont
  {Smith}}]{chanUnitaryprojectiveEntanglementDynamics2019}%
  \BibitemOpen
  \bibfield  {author} {\bibinfo {author} {\bibfnamefont {A.}~\bibnamefont
  {Chan}}, \bibinfo {author} {\bibfnamefont {R.~M.}\ \bibnamefont
  {Nandkishore}}, \bibinfo {author} {\bibfnamefont {M.}~\bibnamefont
  {Pretko}},\ and\ \bibinfo {author} {\bibfnamefont {G.}~\bibnamefont
  {Smith}},\ }\href {https://doi.org/10.1103/PhysRevB.99.224307} {\bibfield
  {journal} {\bibinfo  {journal} {Phys. Rev. B}\ }\textbf {\bibinfo {volume}
  {99}},\ \bibinfo {pages} {224307} (\bibinfo {year} {2019})}\BibitemShut
  {NoStop}%
\bibitem [{\citenamefont {Vasseur}\ \emph {et~al.}(2019)\citenamefont
  {Vasseur}, \citenamefont {Potter}, \citenamefont {You},\ and\ \citenamefont
  {Ludwig}}]{vasseurEntanglementTransitionsHolographic2019}%
  \BibitemOpen
  \bibfield  {author} {\bibinfo {author} {\bibfnamefont {R.}~\bibnamefont
  {Vasseur}}, \bibinfo {author} {\bibfnamefont {A.~C.}\ \bibnamefont {Potter}},
  \bibinfo {author} {\bibfnamefont {Y.-Z.}\ \bibnamefont {You}},\ and\ \bibinfo
  {author} {\bibfnamefont {A.~W.~W.}\ \bibnamefont {Ludwig}},\ }\href
  {https://doi.org/10.1103/PhysRevB.100.134203} {\bibfield  {journal} {\bibinfo
   {journal} {Phys. Rev. B}\ }\textbf {\bibinfo {volume} {100}},\ \bibinfo
  {pages} {134203} (\bibinfo {year} {2019})}\BibitemShut {NoStop}%
\bibitem [{\citenamefont {Zhou}\ and\ \citenamefont
  {Nahum}(2019)}]{zhouEmergentStatisticalMechanics2019}%
  \BibitemOpen
  \bibfield  {author} {\bibinfo {author} {\bibfnamefont {T.}~\bibnamefont
  {Zhou}}\ and\ \bibinfo {author} {\bibfnamefont {A.}~\bibnamefont {Nahum}},\
  }\href {https://doi.org/10.1103/PhysRevB.99.174205} {\bibfield  {journal}
  {\bibinfo  {journal} {Phys. Rev. B}\ }\textbf {\bibinfo {volume} {99}},\
  \bibinfo {pages} {174205} (\bibinfo {year} {2019})}\BibitemShut {NoStop}%
\bibitem [{\citenamefont {Gullans}\ and\ \citenamefont
  {Huse}(2020{\natexlab{a}})}]{gullansScalableProbesMeasurementInduced2020}%
  \BibitemOpen
  \bibfield  {author} {\bibinfo {author} {\bibfnamefont {M.~J.}\ \bibnamefont
  {Gullans}}\ and\ \bibinfo {author} {\bibfnamefont {D.~A.}\ \bibnamefont
  {Huse}},\ }\href {https://doi.org/10.1103/PhysRevLett.125.070606} {\bibfield
  {journal} {\bibinfo  {journal} {Phys. Rev. Lett.}\ }\textbf {\bibinfo
  {volume} {125}},\ \bibinfo {pages} {070606} (\bibinfo {year}
  {2020}{\natexlab{a}})}\BibitemShut {NoStop}%
\bibitem [{\citenamefont {Jian}\ \emph {et~al.}(2020)\citenamefont {Jian},
  \citenamefont {You}, \citenamefont {Vasseur},\ and\ \citenamefont
  {Ludwig}}]{jianMeasurementinducedCriticalityRandom2020}%
  \BibitemOpen
  \bibfield  {author} {\bibinfo {author} {\bibfnamefont {C.-M.}\ \bibnamefont
  {Jian}}, \bibinfo {author} {\bibfnamefont {Y.-Z.}\ \bibnamefont {You}},
  \bibinfo {author} {\bibfnamefont {R.}~\bibnamefont {Vasseur}},\ and\ \bibinfo
  {author} {\bibfnamefont {A.~W.~W.}\ \bibnamefont {Ludwig}},\ }\href
  {https://doi.org/10.1103/PhysRevB.101.104302} {\bibfield  {journal} {\bibinfo
   {journal} {Phys. Rev. B}\ }\textbf {\bibinfo {volume} {101}},\ \bibinfo
  {pages} {104302} (\bibinfo {year} {2020})}\BibitemShut {NoStop}%
\bibitem [{\citenamefont {Fuji}\ and\ \citenamefont
  {Ashida}(2020)}]{fujiMeasurementinducedQuantumCriticality2020}%
  \BibitemOpen
  \bibfield  {author} {\bibinfo {author} {\bibfnamefont {Y.}~\bibnamefont
  {Fuji}}\ and\ \bibinfo {author} {\bibfnamefont {Y.}~\bibnamefont {Ashida}},\
  }\href {https://doi.org/10.1103/PhysRevB.102.054302} {\bibfield  {journal}
  {\bibinfo  {journal} {Phys. Rev. B}\ }\textbf {\bibinfo {volume} {102}},\
  \bibinfo {pages} {054302} (\bibinfo {year} {2020})}\BibitemShut {NoStop}%
\bibitem [{\citenamefont {Zabalo}\ \emph {et~al.}(2020)\citenamefont {Zabalo},
  \citenamefont {Gullans}, \citenamefont {Wilson}, \citenamefont
  {Gopalakrishnan}, \citenamefont {Huse},\ and\ \citenamefont
  {Pixley}}]{zabaloCriticalPropertiesMeasurementinduced2020}%
  \BibitemOpen
  \bibfield  {author} {\bibinfo {author} {\bibfnamefont {A.}~\bibnamefont
  {Zabalo}}, \bibinfo {author} {\bibfnamefont {M.~J.}\ \bibnamefont {Gullans}},
  \bibinfo {author} {\bibfnamefont {J.~H.}\ \bibnamefont {Wilson}}, \bibinfo
  {author} {\bibfnamefont {S.}~\bibnamefont {Gopalakrishnan}}, \bibinfo
  {author} {\bibfnamefont {D.~A.}\ \bibnamefont {Huse}},\ and\ \bibinfo
  {author} {\bibfnamefont {J.~H.}\ \bibnamefont {Pixley}},\ }\href
  {https://doi.org/10.1103/PhysRevB.101.060301} {\bibfield  {journal} {\bibinfo
   {journal} {Phys. Rev. B}\ }\textbf {\bibinfo {volume} {101}},\ \bibinfo
  {pages} {060301} (\bibinfo {year} {2020})}\BibitemShut {NoStop}%
\bibitem [{\citenamefont {Gullans}\ and\ \citenamefont
  {Huse}(2020{\natexlab{b}})}]{gullansDynamicalPurificationPhase2020}%
  \BibitemOpen
  \bibfield  {author} {\bibinfo {author} {\bibfnamefont {M.~J.}\ \bibnamefont
  {Gullans}}\ and\ \bibinfo {author} {\bibfnamefont {D.~A.}\ \bibnamefont
  {Huse}},\ }\href {https://doi.org/10.1103/PhysRevX.10.041020} {\bibfield
  {journal} {\bibinfo  {journal} {Phys. Rev. X}\ }\textbf {\bibinfo {volume}
  {10}},\ \bibinfo {pages} {041020} (\bibinfo {year}
  {2020}{\natexlab{b}})}\BibitemShut {NoStop}%
\bibitem [{\citenamefont {Choi}\ \emph {et~al.}(2020)\citenamefont {Choi},
  \citenamefont {Bao}, \citenamefont {Qi},\ and\ \citenamefont
  {Altman}}]{choiQuantumErrorCorrection2020}%
  \BibitemOpen
  \bibfield  {author} {\bibinfo {author} {\bibfnamefont {S.}~\bibnamefont
  {Choi}}, \bibinfo {author} {\bibfnamefont {Y.}~\bibnamefont {Bao}}, \bibinfo
  {author} {\bibfnamefont {X.-L.}\ \bibnamefont {Qi}},\ and\ \bibinfo {author}
  {\bibfnamefont {E.}~\bibnamefont {Altman}},\ }\href
  {https://doi.org/10.1103/PhysRevLett.125.030505} {\bibfield  {journal}
  {\bibinfo  {journal} {Phys. Rev. Lett.}\ }\textbf {\bibinfo {volume} {125}},\
  \bibinfo {pages} {030505} (\bibinfo {year} {2020})}\BibitemShut {NoStop}%
\bibitem [{\citenamefont {Bao}\ \emph {et~al.}(2020)\citenamefont {Bao},
  \citenamefont {Choi},\ and\ \citenamefont
  {Altman}}]{baoTheoryPhaseTransition2020}%
  \BibitemOpen
  \bibfield  {author} {\bibinfo {author} {\bibfnamefont {Y.}~\bibnamefont
  {Bao}}, \bibinfo {author} {\bibfnamefont {S.}~\bibnamefont {Choi}},\ and\
  \bibinfo {author} {\bibfnamefont {E.}~\bibnamefont {Altman}},\ }\href
  {https://doi.org/10.1103/PhysRevB.101.104301} {\bibfield  {journal} {\bibinfo
   {journal} {Phys. Rev. B}\ }\textbf {\bibinfo {volume} {101}},\ \bibinfo
  {pages} {104301} (\bibinfo {year} {2020})}\BibitemShut {NoStop}%
\bibitem [{\citenamefont {Nahum}\ \emph {et~al.}(2021)\citenamefont {Nahum},
  \citenamefont {Roy}, \citenamefont {Skinner},\ and\ \citenamefont
  {Ruhman}}]{nahumMeasurementEntanglementPhase2021}%
  \BibitemOpen
  \bibfield  {author} {\bibinfo {author} {\bibfnamefont {A.}~\bibnamefont
  {Nahum}}, \bibinfo {author} {\bibfnamefont {S.}~\bibnamefont {Roy}}, \bibinfo
  {author} {\bibfnamefont {B.}~\bibnamefont {Skinner}},\ and\ \bibinfo {author}
  {\bibfnamefont {J.}~\bibnamefont {Ruhman}},\ }\href
  {https://doi.org/10.1103/PRXQuantum.2.010352} {\bibfield  {journal} {\bibinfo
   {journal} {PRX Quantum}\ }\textbf {\bibinfo {volume} {2}},\ \bibinfo {pages}
  {010352} (\bibinfo {year} {2021})}\BibitemShut {NoStop}%
\bibitem [{\citenamefont {Fan}\ \emph {et~al.}(2021)\citenamefont {Fan},
  \citenamefont {Vijay}, \citenamefont {Vishwanath},\ and\ \citenamefont
  {You}}]{PhysRevB.103.174309}%
  \BibitemOpen
  \bibfield  {author} {\bibinfo {author} {\bibfnamefont {R.}~\bibnamefont
  {Fan}}, \bibinfo {author} {\bibfnamefont {S.}~\bibnamefont {Vijay}}, \bibinfo
  {author} {\bibfnamefont {A.}~\bibnamefont {Vishwanath}},\ and\ \bibinfo
  {author} {\bibfnamefont {Y.-Z.}\ \bibnamefont {You}},\ }\href
  {https://doi.org/10.1103/PhysRevB.103.174309} {\bibfield  {journal} {\bibinfo
   {journal} {Phys. Rev. B}\ }\textbf {\bibinfo {volume} {103}},\ \bibinfo
  {pages} {174309} (\bibinfo {year} {2021})}\BibitemShut {NoStop}%
\bibitem [{\citenamefont {Sang}\ and\ \citenamefont
  {Hsieh}(2021)}]{sangMeasurementprotectedQuantumPhases2021}%
  \BibitemOpen
  \bibfield  {author} {\bibinfo {author} {\bibfnamefont {S.}~\bibnamefont
  {Sang}}\ and\ \bibinfo {author} {\bibfnamefont {T.~H.}\ \bibnamefont
  {Hsieh}},\ }\href {https://doi.org/10.1103/PhysRevResearch.3.023200}
  {\bibfield  {journal} {\bibinfo  {journal} {Phys. Rev. Research}\ }\textbf
  {\bibinfo {volume} {3}},\ \bibinfo {pages} {023200} (\bibinfo {year}
  {2021})}\BibitemShut {NoStop}%
\bibitem [{\citenamefont {Alberton}\ \emph {et~al.}(2021)\citenamefont
  {Alberton}, \citenamefont {Buchhold},\ and\ \citenamefont
  {Diehl}}]{albertonEntanglementTransitionMonitored2021}%
  \BibitemOpen
  \bibfield  {author} {\bibinfo {author} {\bibfnamefont {O.}~\bibnamefont
  {Alberton}}, \bibinfo {author} {\bibfnamefont {M.}~\bibnamefont {Buchhold}},\
  and\ \bibinfo {author} {\bibfnamefont {S.}~\bibnamefont {Diehl}},\ }\href
  {https://doi.org/10.1103/PhysRevLett.126.170602} {\bibfield  {journal}
  {\bibinfo  {journal} {Phys. Rev. Lett.}\ }\textbf {\bibinfo {volume} {126}},\
  \bibinfo {pages} {170602} (\bibinfo {year} {2021})}\BibitemShut {NoStop}%
\bibitem [{\citenamefont {Lavasani}\ \emph {et~al.}(2021)\citenamefont
  {Lavasani}, \citenamefont {Alavirad},\ and\ \citenamefont
  {Barkeshli}}]{lavasaniMeasurementinducedTopologicalEntanglement2021}%
  \BibitemOpen
  \bibfield  {author} {\bibinfo {author} {\bibfnamefont {A.}~\bibnamefont
  {Lavasani}}, \bibinfo {author} {\bibfnamefont {Y.}~\bibnamefont {Alavirad}},\
  and\ \bibinfo {author} {\bibfnamefont {M.}~\bibnamefont {Barkeshli}},\ }\href
  {https://doi.org/10.1038/s41567-020-01112-z} {\bibfield  {journal} {\bibinfo
  {journal} {Nat. Phys.}\ }\textbf {\bibinfo {volume} {17}},\ \bibinfo {pages}
  {342} (\bibinfo {year} {2021})}\BibitemShut {NoStop}%
\bibitem [{\citenamefont {Turkeshi}\ \emph {et~al.}(2021)\citenamefont
  {Turkeshi}, \citenamefont {Biella}, \citenamefont {Fazio}, \citenamefont
  {Dalmonte},\ and\ \citenamefont {Schir\'o}}]{PhysRevB.103.224210}%
  \BibitemOpen
  \bibfield  {author} {\bibinfo {author} {\bibfnamefont {X.}~\bibnamefont
  {Turkeshi}}, \bibinfo {author} {\bibfnamefont {A.}~\bibnamefont {Biella}},
  \bibinfo {author} {\bibfnamefont {R.}~\bibnamefont {Fazio}}, \bibinfo
  {author} {\bibfnamefont {M.}~\bibnamefont {Dalmonte}},\ and\ \bibinfo
  {author} {\bibfnamefont {M.}~\bibnamefont {Schir\'o}},\ }\href
  {https://doi.org/10.1103/PhysRevB.103.224210} {\bibfield  {journal} {\bibinfo
   {journal} {Phys. Rev. B}\ }\textbf {\bibinfo {volume} {103}},\ \bibinfo
  {pages} {224210} (\bibinfo {year} {2021})}\BibitemShut {NoStop}%
\bibitem [{\citenamefont {Le~Gal}\ \emph {et~al.}(2022)\citenamefont {Le~Gal},
  \citenamefont {Turkeshi},\ and\ \citenamefont {Schir\`o}}]{LeGal:2022rwf}%
  \BibitemOpen
  \bibfield  {author} {\bibinfo {author} {\bibfnamefont {Y.}~\bibnamefont
  {Le~Gal}}, \bibinfo {author} {\bibfnamefont {X.}~\bibnamefont {Turkeshi}},\
  and\ \bibinfo {author} {\bibfnamefont {M.}~\bibnamefont {Schir\`o}},\
  }\href@noop {} {\  (\bibinfo {year} {2022})},\ \Eprint
  {https://arxiv.org/abs/2210.11937} {arXiv:2210.11937 [cond-mat.stat-mech]}
  \BibitemShut {NoStop}%
\bibitem [{\citenamefont {Jian}\ \emph {et~al.}(2021)\citenamefont {Jian},
  \citenamefont {Liu}, \citenamefont {Chen}, \citenamefont {Swingle},\ and\
  \citenamefont {Zhang}}]{jianMeasurementInducedPhaseTransition2021a}%
  \BibitemOpen
  \bibfield  {author} {\bibinfo {author} {\bibfnamefont {S.-K.}\ \bibnamefont
  {Jian}}, \bibinfo {author} {\bibfnamefont {C.}~\bibnamefont {Liu}}, \bibinfo
  {author} {\bibfnamefont {X.}~\bibnamefont {Chen}}, \bibinfo {author}
  {\bibfnamefont {B.}~\bibnamefont {Swingle}},\ and\ \bibinfo {author}
  {\bibfnamefont {P.}~\bibnamefont {Zhang}},\ }\href
  {https://doi.org/10.1103/PhysRevLett.127.140601} {\bibfield  {journal}
  {\bibinfo  {journal} {Phys. Rev. Lett.}\ }\textbf {\bibinfo {volume} {127}},\
  \bibinfo {pages} {140601} (\bibinfo {year} {2021})}\BibitemShut {NoStop}%
\bibitem [{\citenamefont {Zhang}\ \emph {et~al.}(2022)\citenamefont {Zhang},
  \citenamefont {Liu}, \citenamefont {Jian},\ and\ \citenamefont
  {Chen}}]{zhangUniversalEntanglementTransitions2022}%
  \BibitemOpen
  \bibfield  {author} {\bibinfo {author} {\bibfnamefont {P.}~\bibnamefont
  {Zhang}}, \bibinfo {author} {\bibfnamefont {C.}~\bibnamefont {Liu}}, \bibinfo
  {author} {\bibfnamefont {S.-K.}\ \bibnamefont {Jian}},\ and\ \bibinfo
  {author} {\bibfnamefont {X.}~\bibnamefont {Chen}},\ }\href
  {https://doi.org/10.22331/q-2022-05-27-723} {\bibfield  {journal} {\bibinfo
  {journal} {Quantum}\ }\textbf {\bibinfo {volume} {6}},\ \bibinfo {pages}
  {723} (\bibinfo {year} {2022})}\BibitemShut {NoStop}%
\bibitem [{\citenamefont {Liu}\ \emph {et~al.}(2021)\citenamefont {Liu},
  \citenamefont {Zhang},\ and\ \citenamefont
  {Chen}}]{liuNonunitaryDynamicsSachdevYeKitaev2021}%
  \BibitemOpen
  \bibfield  {author} {\bibinfo {author} {\bibfnamefont {C.}~\bibnamefont
  {Liu}}, \bibinfo {author} {\bibfnamefont {P.}~\bibnamefont {Zhang}},\ and\
  \bibinfo {author} {\bibfnamefont {X.}~\bibnamefont {Chen}},\ }\href
  {https://doi.org/10.21468/SciPostPhys.10.2.048} {\bibfield  {journal}
  {\bibinfo  {journal} {SciPost Phys.}\ }\textbf {\bibinfo {volume} {10}},\
  \bibinfo {pages} {048} (\bibinfo {year} {2021})}\BibitemShut {NoStop}%
\bibitem [{\citenamefont {Zhang}\ \emph {et~al.}(2021)\citenamefont {Zhang},
  \citenamefont {Jian}, \citenamefont {Liu},\ and\ \citenamefont
  {Chen}}]{zhang2021emergent}%
  \BibitemOpen
  \bibfield  {author} {\bibinfo {author} {\bibfnamefont {P.}~\bibnamefont
  {Zhang}}, \bibinfo {author} {\bibfnamefont {S.-K.}\ \bibnamefont {Jian}},
  \bibinfo {author} {\bibfnamefont {C.}~\bibnamefont {Liu}},\ and\ \bibinfo
  {author} {\bibfnamefont {X.}~\bibnamefont {Chen}},\ }\href
  {https://doi.org/10.22331/q-2021-11-16-579} {\bibfield  {journal} {\bibinfo
  {journal} {Quantum}\ }\textbf {\bibinfo {volume} {5}},\ \bibinfo {pages}
  {579} (\bibinfo {year} {2021})}\BibitemShut {NoStop}%
\bibitem [{\citenamefont {Zhang}(2022)}]{zhang2022quantum}%
  \BibitemOpen
  \bibfield  {author} {\bibinfo {author} {\bibfnamefont {P.}~\bibnamefont
  {Zhang}},\ }\href {https://doi.org/10.1007/s11467-022-1162-5} {\bibfield
  {journal} {\bibinfo  {journal} {Front. Phys.}\ }\textbf {\bibinfo {volume}
  {17}},\ \bibinfo {pages} {43201} (\bibinfo {year} {2022})}\BibitemShut
  {NoStop}%
\bibitem [{\citenamefont {Sahu}\ \emph {et~al.}(2022)\citenamefont {Sahu},
  \citenamefont {Jian}, \citenamefont {Bentsen},\ and\ \citenamefont
  {Swingle}}]{PhysRevB.106.224305}%
  \BibitemOpen
  \bibfield  {author} {\bibinfo {author} {\bibfnamefont {S.}~\bibnamefont
  {Sahu}}, \bibinfo {author} {\bibfnamefont {S.-K.}\ \bibnamefont {Jian}},
  \bibinfo {author} {\bibfnamefont {G.}~\bibnamefont {Bentsen}},\ and\ \bibinfo
  {author} {\bibfnamefont {B.}~\bibnamefont {Swingle}},\ }\href
  {https://doi.org/10.1103/PhysRevB.106.224305} {\bibfield  {journal} {\bibinfo
   {journal} {Phys. Rev. B}\ }\textbf {\bibinfo {volume} {106}},\ \bibinfo
  {pages} {224305} (\bibinfo {year} {2022})}\BibitemShut {NoStop}%
\bibitem [{\citenamefont {Rakovszky}\ \emph {et~al.}(2018)\citenamefont
  {Rakovszky}, \citenamefont {Pollmann},\ and\ \citenamefont {{von
  Keyserlingk}}}]{rakovszkyDiffusiveHydrodynamicsOutoftimeordered2018}%
  \BibitemOpen
  \bibfield  {author} {\bibinfo {author} {\bibfnamefont {T.}~\bibnamefont
  {Rakovszky}}, \bibinfo {author} {\bibfnamefont {F.}~\bibnamefont
  {Pollmann}},\ and\ \bibinfo {author} {\bibfnamefont {C.~W.}\ \bibnamefont
  {{von Keyserlingk}}},\ }\href {https://doi.org/10.1103/PhysRevX.8.031058}
  {\bibfield  {journal} {\bibinfo  {journal} {Phys. Rev. X}\ }\textbf {\bibinfo
  {volume} {8}},\ \bibinfo {pages} {031058} (\bibinfo {year}
  {2018})}\BibitemShut {NoStop}%
\bibitem [{\citenamefont {Khemani}\ \emph {et~al.}(2018)\citenamefont
  {Khemani}, \citenamefont {Vishwanath},\ and\ \citenamefont
  {Huse}}]{khemaniOperatorSpreadingEmergence2018a}%
  \BibitemOpen
  \bibfield  {author} {\bibinfo {author} {\bibfnamefont {V.}~\bibnamefont
  {Khemani}}, \bibinfo {author} {\bibfnamefont {A.}~\bibnamefont
  {Vishwanath}},\ and\ \bibinfo {author} {\bibfnamefont {D.~A.}\ \bibnamefont
  {Huse}},\ }\href {https://doi.org/10.1103/PhysRevX.8.031057} {\bibfield
  {journal} {\bibinfo  {journal} {Phys. Rev. X}\ }\textbf {\bibinfo {volume}
  {8}},\ \bibinfo {pages} {031057} (\bibinfo {year} {2018})}\BibitemShut
  {NoStop}%
\bibitem [{\citenamefont {Guo}\ \emph {et~al.}(2019)\citenamefont {Guo},
  \citenamefont {Gu},\ and\ \citenamefont
  {Sachdev}}]{guoTransportChaosLattice2019b}%
  \BibitemOpen
  \bibfield  {author} {\bibinfo {author} {\bibfnamefont {H.}~\bibnamefont
  {Guo}}, \bibinfo {author} {\bibfnamefont {Y.}~\bibnamefont {Gu}},\ and\
  \bibinfo {author} {\bibfnamefont {S.}~\bibnamefont {Sachdev}},\ }\href
  {https://doi.org/10.1103/PhysRevB.100.045140} {\bibfield  {journal} {\bibinfo
   {journal} {Phys. Rev. B}\ }\textbf {\bibinfo {volume} {100}},\ \bibinfo
  {pages} {045140} (\bibinfo {year} {2019})}\BibitemShut {NoStop}%
\bibitem [{\citenamefont {Friedman}\ \emph {et~al.}(2019)\citenamefont
  {Friedman}, \citenamefont {Chan}, \citenamefont {De~Luca},\ and\
  \citenamefont {Chalker}}]{friedmanSpectralStatisticsManyBody2019}%
  \BibitemOpen
  \bibfield  {author} {\bibinfo {author} {\bibfnamefont {A.~J.}\ \bibnamefont
  {Friedman}}, \bibinfo {author} {\bibfnamefont {A.}~\bibnamefont {Chan}},
  \bibinfo {author} {\bibfnamefont {A.}~\bibnamefont {De~Luca}},\ and\ \bibinfo
  {author} {\bibfnamefont {J.~T.}\ \bibnamefont {Chalker}},\ }\href
  {https://doi.org/10.1103/PhysRevLett.123.210603} {\bibfield  {journal}
  {\bibinfo  {journal} {Phys. Rev. Lett.}\ }\textbf {\bibinfo {volume} {123}},\
  \bibinfo {pages} {210603} (\bibinfo {year} {2019})}\BibitemShut {NoStop}%
\bibitem [{\citenamefont {Rakovszky}\ \emph {et~al.}(2019)\citenamefont
  {Rakovszky}, \citenamefont {Pollmann},\ and\ \citenamefont {{von
  Keyserlingk}}}]{rakovszkySubballisticGrowthEnyi2019a}%
  \BibitemOpen
  \bibfield  {author} {\bibinfo {author} {\bibfnamefont {T.}~\bibnamefont
  {Rakovszky}}, \bibinfo {author} {\bibfnamefont {F.}~\bibnamefont
  {Pollmann}},\ and\ \bibinfo {author} {\bibfnamefont {C.~W.}\ \bibnamefont
  {{von Keyserlingk}}},\ }\href
  {https://doi.org/10.1103/PhysRevLett.122.250602} {\bibfield  {journal}
  {\bibinfo  {journal} {Phys. Rev. Lett.}\ }\textbf {\bibinfo {volume} {122}},\
  \bibinfo {pages} {250602} (\bibinfo {year} {2019})}\BibitemShut {NoStop}%
\bibitem [{\citenamefont {Zhou}\ and\ \citenamefont
  {Ludwig}(2020)}]{zhouDiffusiveScalingEnyi2020}%
  \BibitemOpen
  \bibfield  {author} {\bibinfo {author} {\bibfnamefont {T.}~\bibnamefont
  {Zhou}}\ and\ \bibinfo {author} {\bibfnamefont {A.~W.~W.}\ \bibnamefont
  {Ludwig}},\ }\href {https://doi.org/10.1103/PhysRevResearch.2.033020}
  {\bibfield  {journal} {\bibinfo  {journal} {Phys. Rev. Res.}\ }\textbf
  {\bibinfo {volume} {2}},\ \bibinfo {pages} {033020} (\bibinfo {year}
  {2020})}\BibitemShut {NoStop}%
\bibitem [{\citenamefont {Huang}(2020)}]{huangDynamicsRenyiEntanglement2020}%
  \BibitemOpen
  \bibfield  {author} {\bibinfo {author} {\bibfnamefont {Y.}~\bibnamefont
  {Huang}},\ }\href {https://doi.org/10.1088/2633-1357/abd1e2} {\bibfield
  {journal} {\bibinfo  {journal} {IOPSciNotes}\ }\textbf {\bibinfo {volume}
  {1}},\ \bibinfo {pages} {035205} (\bibinfo {year} {2020})}\BibitemShut
  {NoStop}%
\bibitem [{\citenamefont {{Kudler-Flam}}\ \emph {et~al.}(2022)\citenamefont
  {{Kudler-Flam}}, \citenamefont {Sohal},\ and\ \citenamefont
  {Nie}}]{kudler-flamInformationScramblingConservation2022}%
  \BibitemOpen
  \bibfield  {author} {\bibinfo {author} {\bibfnamefont {J.}~\bibnamefont
  {{Kudler-Flam}}}, \bibinfo {author} {\bibfnamefont {R.}~\bibnamefont
  {Sohal}},\ and\ \bibinfo {author} {\bibfnamefont {L.}~\bibnamefont {Nie}},\
  }\href {https://doi.org/10.21468/SciPostPhys.12.4.117} {\bibfield  {journal}
  {\bibinfo  {journal} {SciPost Physics}\ }\textbf {\bibinfo {volume} {12}},\
  \bibinfo {pages} {117} (\bibinfo {year} {2022})}\BibitemShut {NoStop}%
\bibitem [{\citenamefont {Agarwal}\ and\ \citenamefont
  {Xu}(2022)}]{agarwalEmergentSymmetryBrownian2022}%
  \BibitemOpen
  \bibfield  {author} {\bibinfo {author} {\bibfnamefont {L.}~\bibnamefont
  {Agarwal}}\ and\ \bibinfo {author} {\bibfnamefont {S.}~\bibnamefont {Xu}},\
  }\href {https://doi.org/10.1007/JHEP02(2022)045} {\bibfield  {journal}
  {\bibinfo  {journal} {J. High Energ. Phys.}\ }\textbf {\bibinfo {volume}
  {2022}}\bibinfo  {number} { (2)},\ \bibinfo {pages} {45}}\BibitemShut
  {NoStop}%
\bibitem [{\citenamefont
  {Zhang}()}]{zhangInformationScramblingEntanglement2023}%
  \BibitemOpen
\bibfield  {number} {  }\bibfield  {author} {\bibinfo {author} {\bibfnamefont
  {P.}~\bibnamefont {Zhang}},\ }\href@noop {} {\ }\Eprint
  {https://arxiv.org/abs/2301.03189} {arXiv:2301.03189} \BibitemShut {NoStop}%
\bibitem [{\citenamefont {Levitov}\ \emph {et~al.}(1996)\citenamefont
  {Levitov}, \citenamefont {Lee},\ and\ \citenamefont
  {Lesovik}}]{levitovElectronCountingStatistics1996}%
  \BibitemOpen
  \bibfield  {author} {\bibinfo {author} {\bibfnamefont {L.~S.}\ \bibnamefont
  {Levitov}}, \bibinfo {author} {\bibfnamefont {H.}~\bibnamefont {Lee}},\ and\
  \bibinfo {author} {\bibfnamefont {G.~B.}\ \bibnamefont {Lesovik}},\ }\href
  {https://doi.org/10.1063/1.531672} {\bibfield  {journal} {\bibinfo  {journal}
  {J. Math. Phys.}\ }\textbf {\bibinfo {volume} {37}},\ \bibinfo {pages} {4845}
  (\bibinfo {year} {1996})}\BibitemShut {NoStop}%
\bibitem [{\citenamefont {Klich}\ and\ \citenamefont
  {Levitov}(2009)}]{klichQuantumNoiseEntanglement2009}%
  \BibitemOpen
  \bibfield  {author} {\bibinfo {author} {\bibfnamefont {I.}~\bibnamefont
  {Klich}}\ and\ \bibinfo {author} {\bibfnamefont {L.}~\bibnamefont
  {Levitov}},\ }\href {https://doi.org/10.1103/PhysRevLett.102.100502}
  {\bibfield  {journal} {\bibinfo  {journal} {Phys. Rev. Lett.}\ }\textbf
  {\bibinfo {volume} {102}},\ \bibinfo {pages} {100502} (\bibinfo {year}
  {2009})}\BibitemShut {NoStop}%
\bibitem [{\citenamefont {Song}\ \emph {et~al.}(2011)\citenamefont {Song},
  \citenamefont {Flindt}, \citenamefont {Rachel}, \citenamefont {Klich},\ and\
  \citenamefont {Le~Hur}}]{songEntanglementEntropyCharge2011}%
  \BibitemOpen
  \bibfield  {author} {\bibinfo {author} {\bibfnamefont {H.~F.}\ \bibnamefont
  {Song}}, \bibinfo {author} {\bibfnamefont {C.}~\bibnamefont {Flindt}},
  \bibinfo {author} {\bibfnamefont {S.}~\bibnamefont {Rachel}}, \bibinfo
  {author} {\bibfnamefont {I.}~\bibnamefont {Klich}},\ and\ \bibinfo {author}
  {\bibfnamefont {K.}~\bibnamefont {Le~Hur}},\ }\href
  {https://doi.org/10.1103/PhysRevB.83.161408} {\bibfield  {journal} {\bibinfo
  {journal} {Phys. Rev. B}\ }\textbf {\bibinfo {volume} {83}},\ \bibinfo
  {pages} {161408} (\bibinfo {year} {2011})}\BibitemShut {NoStop}%
\bibitem [{\citenamefont {Calabrese}\ \emph {et~al.}(2012)\citenamefont
  {Calabrese}, \citenamefont {Mintchev},\ and\ \citenamefont
  {Vicari}}]{calabreseExactRelationsParticle2012}%
  \BibitemOpen
  \bibfield  {author} {\bibinfo {author} {\bibfnamefont {P.}~\bibnamefont
  {Calabrese}}, \bibinfo {author} {\bibfnamefont {M.}~\bibnamefont
  {Mintchev}},\ and\ \bibinfo {author} {\bibfnamefont {E.}~\bibnamefont
  {Vicari}},\ }\href {https://doi.org/10.1209/0295-5075/98/20003} {\bibfield
  {journal} {\bibinfo  {journal} {EPL}\ }\textbf {\bibinfo {volume} {98}},\
  \bibinfo {pages} {20003} (\bibinfo {year} {2012})}\BibitemShut {NoStop}%
\bibitem [{\citenamefont {S{\"u}sstrunk}\ and\ \citenamefont
  {Ivanov}(2013)}]{susstrunkFreeFermionsLine2013}%
  \BibitemOpen
  \bibfield  {author} {\bibinfo {author} {\bibfnamefont {R.}~\bibnamefont
  {S{\"u}sstrunk}}\ and\ \bibinfo {author} {\bibfnamefont {D.~A.}\ \bibnamefont
  {Ivanov}},\ }\href {https://doi.org/10.1209/0295-5075/100/60009} {\bibfield
  {journal} {\bibinfo  {journal} {EPL}\ }\textbf {\bibinfo {volume} {100}},\
  \bibinfo {pages} {60009} (\bibinfo {year} {2013})}\BibitemShut {NoStop}%
\bibitem [{\citenamefont {Song}\ \emph {et~al.}(2010)\citenamefont {Song},
  \citenamefont {Rachel},\ and\ \citenamefont
  {Le~Hur}}]{songGeneralRelationEntanglement2010}%
  \BibitemOpen
  \bibfield  {author} {\bibinfo {author} {\bibfnamefont {H.~F.}\ \bibnamefont
  {Song}}, \bibinfo {author} {\bibfnamefont {S.}~\bibnamefont {Rachel}},\ and\
  \bibinfo {author} {\bibfnamefont {K.}~\bibnamefont {Le~Hur}},\ }\href
  {https://doi.org/10.1103/PhysRevB.82.012405} {\bibfield  {journal} {\bibinfo
  {journal} {Phys. Rev. B}\ }\textbf {\bibinfo {volume} {82}},\ \bibinfo
  {pages} {012405} (\bibinfo {year} {2010})}\BibitemShut {NoStop}%
\bibitem [{\citenamefont {Levine}\ \emph {et~al.}(2012)\citenamefont {Levine},
  \citenamefont {Bantegui},\ and\ \citenamefont
  {Burg}}]{levineFullCountingStatistics2012a}%
  \BibitemOpen
  \bibfield  {author} {\bibinfo {author} {\bibfnamefont {G.~C.}\ \bibnamefont
  {Levine}}, \bibinfo {author} {\bibfnamefont {M.~J.}\ \bibnamefont
  {Bantegui}},\ and\ \bibinfo {author} {\bibfnamefont {J.~A.}\ \bibnamefont
  {Burg}},\ }\href {https://doi.org/10.1103/PhysRevB.86.174202} {\bibfield
  {journal} {\bibinfo  {journal} {Phys. Rev. B}\ }\textbf {\bibinfo {volume}
  {86}},\ \bibinfo {pages} {174202} (\bibinfo {year} {2012})}\BibitemShut
  {NoStop}%
\bibitem [{\citenamefont {Song}\ \emph {et~al.}(2012)\citenamefont {Song},
  \citenamefont {Rachel}, \citenamefont {Flindt}, \citenamefont {Klich},
  \citenamefont {Laflorencie},\ and\ \citenamefont
  {Le~Hur}}]{songBipartiteFluctuationsProbe2012}%
  \BibitemOpen
  \bibfield  {author} {\bibinfo {author} {\bibfnamefont {H.~F.}\ \bibnamefont
  {Song}}, \bibinfo {author} {\bibfnamefont {S.}~\bibnamefont {Rachel}},
  \bibinfo {author} {\bibfnamefont {C.}~\bibnamefont {Flindt}}, \bibinfo
  {author} {\bibfnamefont {I.}~\bibnamefont {Klich}}, \bibinfo {author}
  {\bibfnamefont {N.}~\bibnamefont {Laflorencie}},\ and\ \bibinfo {author}
  {\bibfnamefont {K.}~\bibnamefont {Le~Hur}},\ }\href
  {https://doi.org/10.1103/PhysRevB.85.035409} {\bibfield  {journal} {\bibinfo
  {journal} {Phys. Rev. B}\ }\textbf {\bibinfo {volume} {85}},\ \bibinfo
  {pages} {035409} (\bibinfo {year} {2012})}\BibitemShut {NoStop}%
\bibitem [{\citenamefont {Ivanov}\ and\ \citenamefont
  {Abanov}(2013)}]{ivanovCharacterizingCorrelationsFull2013}%
  \BibitemOpen
  \bibfield  {author} {\bibinfo {author} {\bibfnamefont {D.~A.}\ \bibnamefont
  {Ivanov}}\ and\ \bibinfo {author} {\bibfnamefont {A.~G.}\ \bibnamefont
  {Abanov}},\ }\href {https://doi.org/10.1103/PhysRevE.87.022114} {\bibfield
  {journal} {\bibinfo  {journal} {Phys. Rev. E}\ }\textbf {\bibinfo {volume}
  {87}},\ \bibinfo {pages} {022114} (\bibinfo {year} {2013})}\BibitemShut
  {NoStop}%
\bibitem [{\citenamefont {Eisler}(2013)}]{eislerUniversalityFullCounting2013}%
  \BibitemOpen
  \bibfield  {author} {\bibinfo {author} {\bibfnamefont {V.}~\bibnamefont
  {Eisler}},\ }\href {https://doi.org/10.1103/PhysRevLett.111.080402}
  {\bibfield  {journal} {\bibinfo  {journal} {Phys. Rev. Lett.}\ }\textbf
  {\bibinfo {volume} {111}},\ \bibinfo {pages} {080402} (\bibinfo {year}
  {2013})}\BibitemShut {NoStop}%
\bibitem [{\citenamefont {Eisler}\ and\ \citenamefont
  {R{\'a}cz}(2013)}]{eislerFullCountingStatistics2013}%
  \BibitemOpen
  \bibfield  {author} {\bibinfo {author} {\bibfnamefont {V.}~\bibnamefont
  {Eisler}}\ and\ \bibinfo {author} {\bibfnamefont {Z.}~\bibnamefont
  {R{\'a}cz}},\ }\href {https://doi.org/10.1103/PhysRevLett.110.060602}
  {\bibfield  {journal} {\bibinfo  {journal} {Phys. Rev. Lett.}\ }\textbf
  {\bibinfo {volume} {110}},\ \bibinfo {pages} {060602} (\bibinfo {year}
  {2013})}\BibitemShut {NoStop}%
\bibitem [{\citenamefont {Klich}(2014)}]{klichNoteFullCounting2014}%
  \BibitemOpen
  \bibfield  {author} {\bibinfo {author} {\bibfnamefont {I.}~\bibnamefont
  {Klich}},\ }\href {https://doi.org/10.1088/1742-5468/2014/11/P11006}
  {\bibfield  {journal} {\bibinfo  {journal} {J. Stat. Mech.}\ }\textbf
  {\bibinfo {volume} {2014}},\ \bibinfo {pages} {P11006} (\bibinfo {year}
  {2014})}\BibitemShut {NoStop}%
\bibitem [{\citenamefont {Bertini}\ \emph {et~al.}()\citenamefont {Bertini},
  \citenamefont {Calabrese}, \citenamefont {Collura}, \citenamefont {Klobas},\
  and\ \citenamefont {Rylands}}]{bertiniNonequilibriumFullCounting2022}%
  \BibitemOpen
  \bibfield  {author} {\bibinfo {author} {\bibfnamefont {B.}~\bibnamefont
  {Bertini}}, \bibinfo {author} {\bibfnamefont {P.}~\bibnamefont {Calabrese}},
  \bibinfo {author} {\bibfnamefont {M.}~\bibnamefont {Collura}}, \bibinfo
  {author} {\bibfnamefont {K.}~\bibnamefont {Klobas}},\ and\ \bibinfo {author}
  {\bibfnamefont {C.}~\bibnamefont {Rylands}},\ }\href@noop {} {\ }\Eprint
  {https://arxiv.org/abs/2212.06188} {arXiv:2212.06188} \BibitemShut {NoStop}%
\bibitem [{\citenamefont {Barratt}\ \emph {et~al.}(2022)\citenamefont
  {Barratt}, \citenamefont {Agrawal}, \citenamefont {Gopalakrishnan},
  \citenamefont {Huse}, \citenamefont {Vasseur},\ and\ \citenamefont
  {Potter}}]{PotterFieldTheoryCharge2022}%
  \BibitemOpen
  \bibfield  {author} {\bibinfo {author} {\bibfnamefont {F.}~\bibnamefont
  {Barratt}}, \bibinfo {author} {\bibfnamefont {U.}~\bibnamefont {Agrawal}},
  \bibinfo {author} {\bibfnamefont {S.}~\bibnamefont {Gopalakrishnan}},
  \bibinfo {author} {\bibfnamefont {D.~A.}\ \bibnamefont {Huse}}, \bibinfo
  {author} {\bibfnamefont {R.}~\bibnamefont {Vasseur}},\ and\ \bibinfo {author}
  {\bibfnamefont {A.~C.}\ \bibnamefont {Potter}},\ }\href
  {https://doi.org/10.1103/PhysRevLett.129.120604} {\bibfield  {journal}
  {\bibinfo  {journal} {Phys. Rev. Lett.}\ }\textbf {\bibinfo {volume} {129}},\
  \bibinfo {pages} {120604} (\bibinfo {year} {2022})}\BibitemShut {NoStop}%
\bibitem [{\citenamefont {McCulloch}\ \emph {et~al.}()\citenamefont
  {McCulloch}, \citenamefont {De~Nardis}, \citenamefont {Gopalakrishnan},\ and\
  \citenamefont {Vasseur}}]{mccullochFullCountingStatistics2023}%
  \BibitemOpen
  \bibfield  {author} {\bibinfo {author} {\bibfnamefont {E.}~\bibnamefont
  {McCulloch}}, \bibinfo {author} {\bibfnamefont {J.}~\bibnamefont
  {De~Nardis}}, \bibinfo {author} {\bibfnamefont {S.}~\bibnamefont
  {Gopalakrishnan}},\ and\ \bibinfo {author} {\bibfnamefont {R.}~\bibnamefont
  {Vasseur}},\ }\href@noop {} {\ }\Eprint {https://arxiv.org/abs/2302.01355}
  {arXiv:2302.01355} \BibitemShut {NoStop}%
\bibitem [{\citenamefont {Oshima}\ and\ \citenamefont
  {Fuji}(2023)}]{FujiChargeFluctuationChargeresolved2023}%
  \BibitemOpen
  \bibfield  {author} {\bibinfo {author} {\bibfnamefont {H.}~\bibnamefont
  {Oshima}}\ and\ \bibinfo {author} {\bibfnamefont {Y.}~\bibnamefont {Fuji}},\
  }\href {https://doi.org/10.1103/PhysRevB.107.014308} {\bibfield  {journal}
  {\bibinfo  {journal} {Phys. Rev. B}\ }\textbf {\bibinfo {volume} {107}},\
  \bibinfo {pages} {014308} (\bibinfo {year} {2023})}\BibitemShut {NoStop}%
\bibitem [{\citenamefont {Chen}\ \emph {et~al.}(2022)\citenamefont {Chen},
  \citenamefont {Tu}, \citenamefont {Meng},\ and\ \citenamefont
  {Cheng}}]{PhysRevB.106.094415}%
  \BibitemOpen
  \bibfield  {author} {\bibinfo {author} {\bibfnamefont {B.-B.}\ \bibnamefont
  {Chen}}, \bibinfo {author} {\bibfnamefont {H.-H.}\ \bibnamefont {Tu}},
  \bibinfo {author} {\bibfnamefont {Z.~Y.}\ \bibnamefont {Meng}},\ and\
  \bibinfo {author} {\bibfnamefont {M.}~\bibnamefont {Cheng}},\ }\href
  {https://doi.org/10.1103/PhysRevB.106.094415} {\bibfield  {journal} {\bibinfo
   {journal} {Phys. Rev. B}\ }\textbf {\bibinfo {volume} {106}},\ \bibinfo
  {pages} {094415} (\bibinfo {year} {2022})}\BibitemShut {NoStop}%
\bibitem [{\citenamefont {Wang}\ \emph {et~al.}(2021)\citenamefont {Wang},
  \citenamefont {Cheng},\ and\ \citenamefont {Meng}}]{PhysRevB.104.L081109}%
  \BibitemOpen
  \bibfield  {author} {\bibinfo {author} {\bibfnamefont {Y.-C.}\ \bibnamefont
  {Wang}}, \bibinfo {author} {\bibfnamefont {M.}~\bibnamefont {Cheng}},\ and\
  \bibinfo {author} {\bibfnamefont {Z.~Y.}\ \bibnamefont {Meng}},\ }\href
  {https://doi.org/10.1103/PhysRevB.104.L081109} {\bibfield  {journal}
  {\bibinfo  {journal} {Phys. Rev. B}\ }\textbf {\bibinfo {volume} {104}},\
  \bibinfo {pages} {L081109} (\bibinfo {year} {2021})}\BibitemShut {NoStop}%
\bibitem [{\citenamefont {Wang}\ \emph {et~al.}(2022)\citenamefont {Wang},
  \citenamefont {Ma}, \citenamefont {Cheng},\ and\ \citenamefont
  {Meng}}]{10.21468/SciPostPhys.13.6.123}%
  \BibitemOpen
  \bibfield  {author} {\bibinfo {author} {\bibfnamefont {Y.-C.}\ \bibnamefont
  {Wang}}, \bibinfo {author} {\bibfnamefont {N.}~\bibnamefont {Ma}}, \bibinfo
  {author} {\bibfnamefont {M.}~\bibnamefont {Cheng}},\ and\ \bibinfo {author}
  {\bibfnamefont {Z.~Y.}\ \bibnamefont {Meng}},\ }\href
  {https://doi.org/10.21468/SciPostPhys.13.6.123} {\bibfield  {journal}
  {\bibinfo  {journal} {SciPost Phys.}\ }\textbf {\bibinfo {volume} {13}},\
  \bibinfo {pages} {123} (\bibinfo {year} {2022})}\BibitemShut {NoStop}%
\bibitem [{\citenamefont {Israel}(1976)}]{Israel:1976ur}%
  \BibitemOpen
  \bibfield  {author} {\bibinfo {author} {\bibfnamefont {W.}~\bibnamefont
  {Israel}},\ }\href {https://doi.org/10.1016/0375-9601(76)90178-X} {\bibfield
  {journal} {\bibinfo  {journal} {Phys. Lett. A}\ }\textbf {\bibinfo {volume}
  {57}},\ \bibinfo {pages} {107} (\bibinfo {year} {1976})}\BibitemShut
  {NoStop}%
\bibitem [{\citenamefont {Maldacena}(2003)}]{maldacenaEternalBlackHoles2003}%
  \BibitemOpen
  \bibfield  {author} {\bibinfo {author} {\bibfnamefont {J.}~\bibnamefont
  {Maldacena}},\ }\href {https://doi.org/10.1088/1126-6708/2003/04/021}
  {\bibfield  {journal} {\bibinfo  {journal} {J. High Energy Phys.}\ }\textbf
  {\bibinfo {volume} {2003}}\bibinfo  {number} { (04)},\ \bibinfo {pages}
  {021}}\BibitemShut {NoStop}%
\bibitem [{\citenamefont {Kitaev}\ and\ \citenamefont
  {Suh}(2018)}]{Kitaev:2017awl}%
  \BibitemOpen
\bibfield  {number} {  }\bibfield  {author} {\bibinfo {author} {\bibfnamefont
  {A.}~\bibnamefont {Kitaev}}\ and\ \bibinfo {author} {\bibfnamefont {S.~J.}\
  \bibnamefont {Suh}},\ }\href {https://doi.org/10.1007/JHEP05(2018)183}
  {\bibfield  {journal} {\bibinfo  {journal} {JHEP}\ }\textbf {\bibinfo
  {volume} {05}},\ \bibinfo {pages} {183}}\BibitemShut {NoStop}%
\bibitem [{\citenamefont {Gu}\ \emph {et~al.}(2020)\citenamefont {Gu},
  \citenamefont {Kitaev}, \citenamefont {Sachdev},\ and\ \citenamefont
  {Tarnopolsky}}]{Gu:2019jub}%
  \BibitemOpen
  \bibfield  {author} {\bibinfo {author} {\bibfnamefont {Y.}~\bibnamefont
  {Gu}}, \bibinfo {author} {\bibfnamefont {A.}~\bibnamefont {Kitaev}}, \bibinfo
  {author} {\bibfnamefont {S.}~\bibnamefont {Sachdev}},\ and\ \bibinfo {author}
  {\bibfnamefont {G.}~\bibnamefont {Tarnopolsky}},\ }\href
  {https://doi.org/10.1007/JHEP02(2020)157} {\bibfield  {journal} {\bibinfo
  {journal} {JHEP}\ }\textbf {\bibinfo {volume} {02}},\ \bibinfo {pages}
  {157}}\BibitemShut {NoStop}%
\bibitem [{\citenamefont {Maldacena}\ and\ \citenamefont
  {Stanford}(2016)}]{maldacena2016remarks}%
  \BibitemOpen
  \bibfield  {author} {\bibinfo {author} {\bibfnamefont {J.}~\bibnamefont
  {Maldacena}}\ and\ \bibinfo {author} {\bibfnamefont {D.}~\bibnamefont
  {Stanford}},\ }\href {https://doi.org/10.1103/PhysRevD.94.106002} {\bibfield
  {journal} {\bibinfo  {journal} {Phys. Rev. D}\ }\textbf {\bibinfo {volume}
  {94}},\ \bibinfo {pages} {106002} (\bibinfo {year} {2016})}\BibitemShut
  {NoStop}%
\bibitem [{\citenamefont {Zhai}(2021)}]{zhai_2021}%
  \BibitemOpen
  \bibfield  {author} {\bibinfo {author} {\bibfnamefont {H.}~\bibnamefont
  {Zhai}},\ }\href {https://doi.org/10.1017/9781108595216} {\emph {\bibinfo
  {title} {Ultracold Atomic Physics}}}\ (\bibinfo  {publisher} {Cambridge
  University Press},\ \bibinfo {year} {2021})\BibitemShut {NoStop}%
\bibitem [{\citenamefont {Zhang}(2020)}]{zhang2020entanglement}%
  \BibitemOpen
  \bibfield  {author} {\bibinfo {author} {\bibfnamefont {P.}~\bibnamefont
  {Zhang}},\ }\href {https://doi.org/10.1007/JHEP06(2020)143} {\bibfield
  {journal} {\bibinfo  {journal} {J. High Energ. Phys.}\ }\textbf {\bibinfo
  {volume} {2020}}\bibinfo  {number} { (6)},\ \bibinfo {pages}
  {143}}\BibitemShut {NoStop}%
\bibitem [{\citenamefont {Tirrito}\ \emph {et~al.}()\citenamefont {Tirrito},
  \citenamefont {Santini}, \citenamefont {Fazio},\ and\ \citenamefont
  {Collura}}]{tirritoFullCountingStatistics2023}%
  \BibitemOpen
\bibfield  {number} {  }\bibfield  {author} {\bibinfo {author} {\bibfnamefont
  {E.}~\bibnamefont {Tirrito}}, \bibinfo {author} {\bibfnamefont
  {A.}~\bibnamefont {Santini}}, \bibinfo {author} {\bibfnamefont
  {R.}~\bibnamefont {Fazio}},\ and\ \bibinfo {author} {\bibfnamefont
  {M.}~\bibnamefont {Collura}},\ }\href@noop {} {\ }\Eprint
  {https://arxiv.org/abs/2212.09405} {arXiv:2212.09405} \BibitemShut {NoStop}%
\end{thebibliography}%
\end{document}